\documentclass[12pt,preprint]{aastex}
\def\p/{\mbox{$^1$}}
\def\pp/{\mbox{$^2$}}
\def\ppp/{\mbox{$^3$}}
\def\pppp/{\mbox{$^4$}}
\def\m/{\mbox{$^{-1}$}}
\def\mm/{\mbox{$^{-2}$}}
\def\mmm/{\mbox{$^{-3}$}}
\def\mmmm/{\mbox{$^{-4}$}}
\def\Ms/{\mbox{M$_\odot$}}
\def\ebv{\mbox{$E(4405-5495)$}}
\def\rv{\mbox{$R_{5495}$}}
\def\chimin{\mbox{$\chi^2_{\rm min}$}}

\slugcomment{To appear in the September 2004 issue of Publications of 
the Astronomical Society of the Pacific}
\begin{document}

\title{CHORIZOS: a CHi-square cOde for parameteRized modelIng and 
       characteriZation of phOtometry and Spectrophotometry}
\shorttitle{CHORIZOS: a code for (spectro)photometry modeling}

\author{Jes\'us Ma\'{\i}z-Apell\'aniz\altaffilmark{1}}
\affil{Space Telescope Science Institute\altaffilmark{2}, 3700 San Martin 
Drive, Baltimore, MD 21218, U.S.A.}
\email{jmaiz@stsci.edu}


\altaffiltext{1}{Affiliated with the Space Telescope Division of the European 
Space Agency, ESTEC, Noordwijk, Netherlands.}
\altaffiltext{2}{The Space Telescope Science Institute is operated by the
Association of Universities for Research in Astronomy, Inc. under NASA
contract No. NAS5-26555.}

\begin{abstract}
We have developed a CHi-square cOde for parameteRized modelIng and 
characteriZation of phOtometry and Spectrophotometry (CHORIZOS).
CHORIZOS can use up to two intrinsic free parameters (e.g. temperature and 
gravity for stars; type and redshift for galaxies; or age and metallicity for
stellar clusters) and two extrinsic ones (amount and type of extinction).
The code uses $\chi^2$ minimization to find all models compatible with
the observed data in the model $N$-dimensional ($N=1,2,3,4$) parameter space. CHORIZOS
can use either correlated or uncorrelated colors as input and is especially
designed to identify possible parameter degeneracies and multiple solutions.
The code is written in IDL and is available to the astronomical 
community. Here we present the techniques used, test the 
code, apply it to a few well-known astronomical problems, and suggest possible 
applications. As a first scientific result from CHORIZOS, we confirm from photometry 
the need for a revised temperature-spectral type scale for OB stars previously
derived from spectroscopy.
\end{abstract}


\section{Introduction}

	A general problem in astronomy is that of finding the correspondence 
between the observed spectrophotometric and/or photometric properties and a 
series of models parameterized in terms of physical quantities. Perhaps the 
best well-known example is the utilization of Johnson $U-B$ and $B-V$ colors to 
measure the temperature and extinction of main-sequence stars (see, e.g. 
\citealt{BinnMerr98}) but the problem appears in a variety of contexts, such as 
the calculation of photometric redshifts with optical/IR photometry 
\citep{Koo99,Beni00}, of extinction laws using a combination of UV-to-IR
spectroscopy and photometry \citep{Cardetal89,Fitz99}, or of stellar cluster 
ages and metallicities using broad-band colors 
\citep{Gira00,Whitetal99b,deGretal03}. Each one of those cases has its
own specific peculiarities, but they can all be considered as examples of the
following general problem. A family of spectral energy distribution (SED) models
$f(\lambda;p_1,p_2\ldots,p_N)$ is generated as a function of $N$ parameters. 
Some
of those parameters $p_i$ may depend on the nature and distance to the objects 
(e.g.  temperature, age, metallicity, or redshift) while others depend on the 
properties of the intervening ISM (amount and type of extinction). The first 
type of parameters will be called in this paper intrinsic (to the object) and the 
second type extrinsic (to the object).
Our data consists of one or several objects with measured
magnitudes $m_1,m_2\ldots m_{M+1}$ ($M\ge N$), each with its independent 
uncertainty $s_1,s_2\ldots s_{M+1}$ from which we can derive $M$ independent 
colors\footnote{Alternatively, we can have measured colors with their
corresponding uncertainties.} $c_1,c_2\ldots c_{M}$.
The general problem can be then expressed as finding what model SEDs are 
compatible with the observed colors\footnote{The photometric redshift case
requires a slight reformulation of the problem because one of the parameters,
redshift, changes not only colors but also magnitudes.}. The solution could
be:

\begin{itemize}
  \item Unique, if the observed values for the colors and their uncertainties 
	determine a single set of connected SED models (i.e. a connected 
        $N-$volume in $N-$dimensional parameter space) that can be described 
	by a single set of parameters with their corresponding uncertainties.
  \item Multiple, if the provided colors do not allow to differentiate
	between two or more sets of connected SED models.
  \item Non-existent, if the properties of the object fall outside the 
	parameter range of the SED models, the chosen SED models are an 
	incorrect description of the object, or the photometry has
	systematic errors.
\end{itemize}

	Several related codes are discussed in the literature (see e.g. 
\citealt{Romaetal02,Beni00,deGretal03}) but they all have one shortcoming:
they are designed to deal only with an specific case of the general problem
(stellar temperatures, photometric redhifts, cluster ages). Furthermore, some
of the codes are discussed in an article but are not available to the
astronomical community, thus hampering the testing of results by other groups.
Finally, the treatment of how uncertainties in the measured quantities affect
the derived parameters is in many cases poor or inexistent (some photometric 
redshift codes are an exception, see e.g. \citealt{Beni00}). 

	We discuss in this paper CHORIZOS, a code that solves the general problem
of identifying which models are compatible with an observed set of colors. In
section 2 we analyze the problems that have to be dealt with and in section 3
we discuss the techniques to overcome them. In section 4 we apply an
implementation of the algorithm to a number of cases and we end in section 5 
with a summary.

\section{Description}

\subsection{Problem definition}

	We start by defining two different cases as a function of the number
of colors and parameters, $M=N$ and $M>N$. For $M=N$, we can establish as many
equations (sets of model parameters that produce a given color) as 
unknowns (model parameters) and the problem can be treated as that of finding
an exact solution to a (complicated) algebraic system of equations. Barring strict
degeneracies in the system of equations, 
one can find either zero, one, or a 
finite number of solutions in $N-$dimensional parameter space depending on the 
specific topology of the $N-$volume defined in $M-$dimensional color space by 
all the possible parameter combinations. Adding the corresponding 
uncertainties associated with each color produces at least one connected 
$N-$volume around each of the solutions in parameter space and can also 
generate new $N-$volumes disconnected from any of the strict solutions and/or
establish connections between solutions\footnote{Strictly speaking, once
uncertainties are included, any solution is possible. In practice, we normally
consider that the probability of a color having a real value many sigmas away 
from its measured one is zero.}. For $M>N$, the problem has more equations than
unknowns and no exact solutions should be expected. However, approximate
solutions that are compatible with the measured uncertainties can still be found. 
In this case, our goal should be to find that $N-$volume of approximate 
solutions by using e.g $\chi^2$ minimization.

\subsection{$M=N$: the main-sequence $B-V$ vs. $U-B$ example}

\begin{figure}
\centerline{\includegraphics*[width=\linewidth]{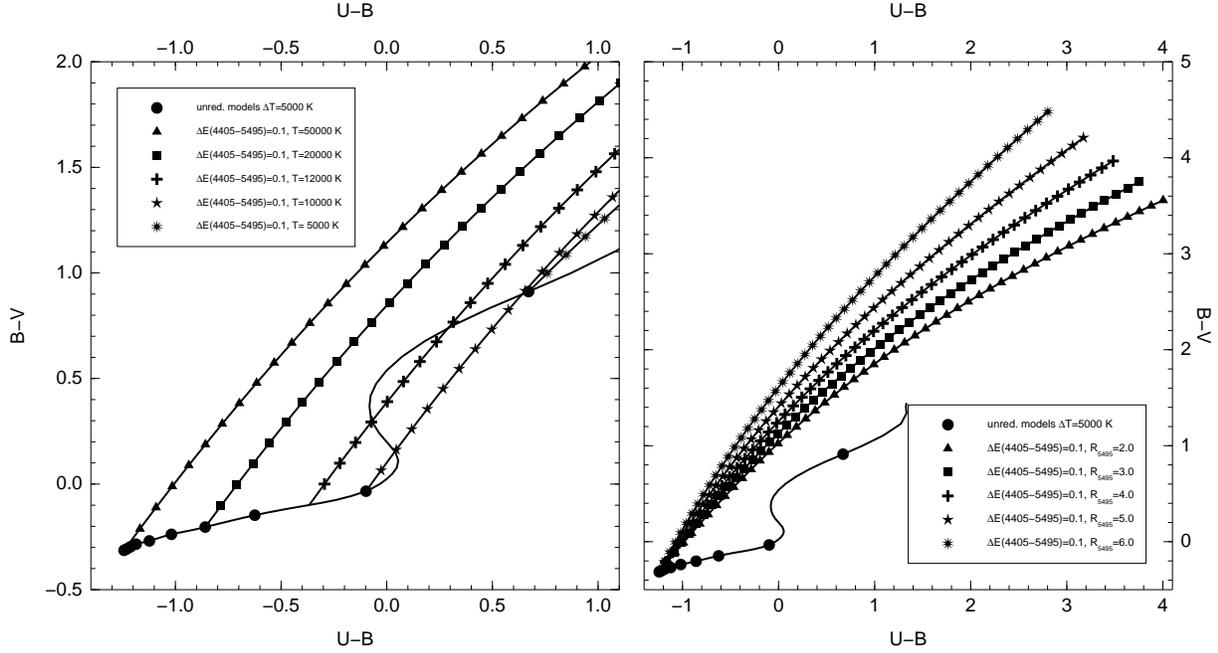}}
\caption{(left) $U-B$ vs. $B-V$ color-color plot for $Z=0.0$, main-sequence
Kurucz atmospheric models. The line with circles indicates the location of the
unredenned values as a function of temperature, starting at $T=50\,000$ K, with
the circles marking those points where the temperature is a multiple of 5\,000 K.
The rest of the lines indicate the colors as a function of reddening using the
\citet{Cardetal89} law with $R_{5495}=3.1$ for five different temperatures. 
Symbols are plotted at intervals of $\Delta\ebv\ = 0.1$0. (right) Same as left 
but now plotting colors as a function of reddening for five different values of 
$R_{5495}$ using the same family of extinction laws for a temperature of 
50\,000 K. The scale has been enlarged to include all unredenned models with 
$T=3\,500-50\,000$ K and all 50\,000 K models up to $\ebv = 5.0$.}
\label{ubvcolorcolor1}
\end{figure}

\begin{figure}
\centerline{\includegraphics*[width=\linewidth]{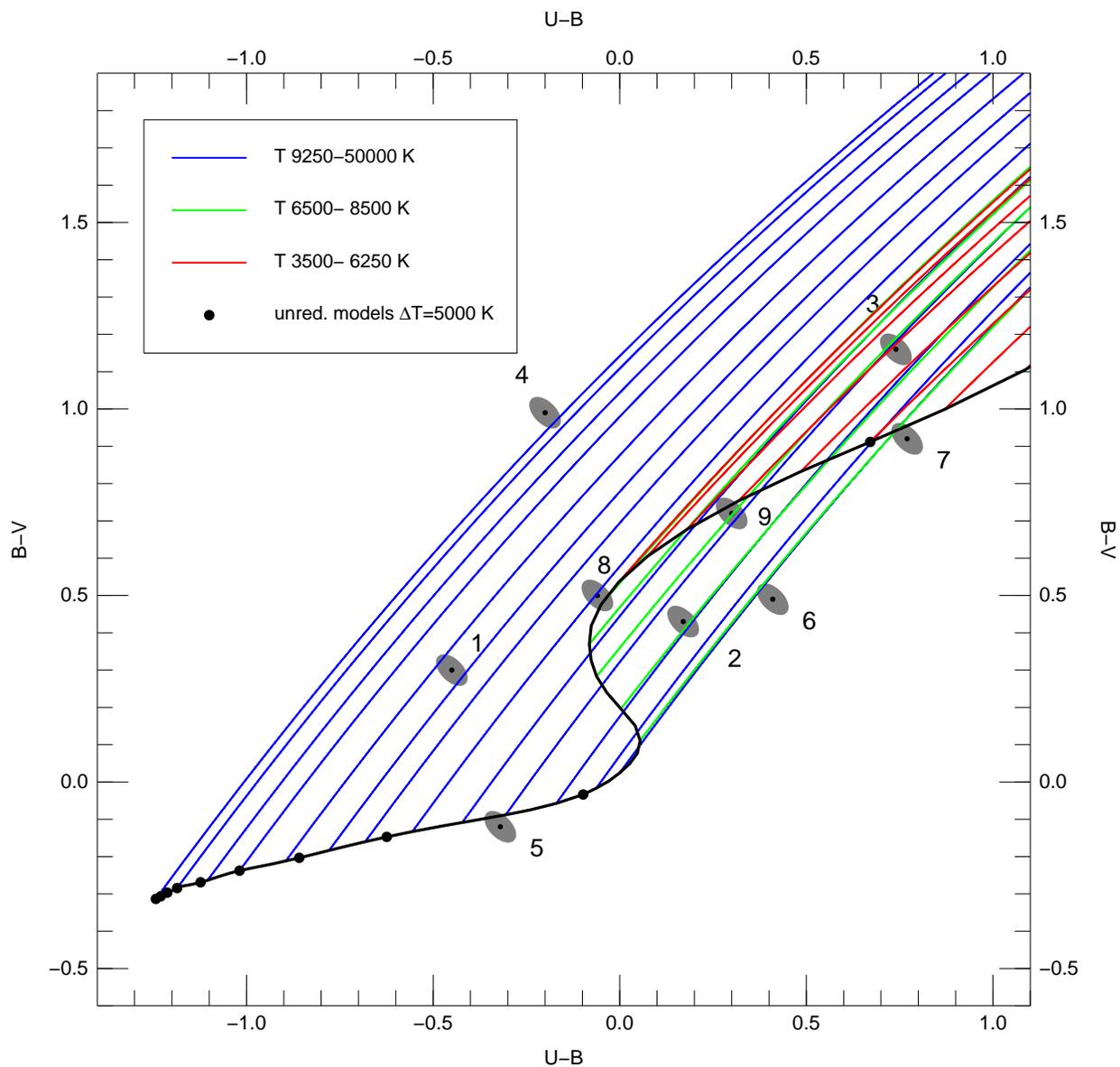}}
\caption{$B-V$ vs. $U-B$ color-color plot for the same conditions as in 
Fig.~\ref{ubvcolorcolor1}. Here we include a higher number of 
extinguished models using a color
code to differentiate among temperature ranges which are relevant to determine
the number of possible temperature + reddening solutions for a given $(U-B)$ 
+ $(B-V)$ color combination. Nine examples are marked, each one of them with
$\sigma_U = \sigma_B = \sigma_V = 0.026$.}
\label{ubvcolorcolor2}
\end{figure}

\begin{figure}
\centerline{\includegraphics*[width=\linewidth]{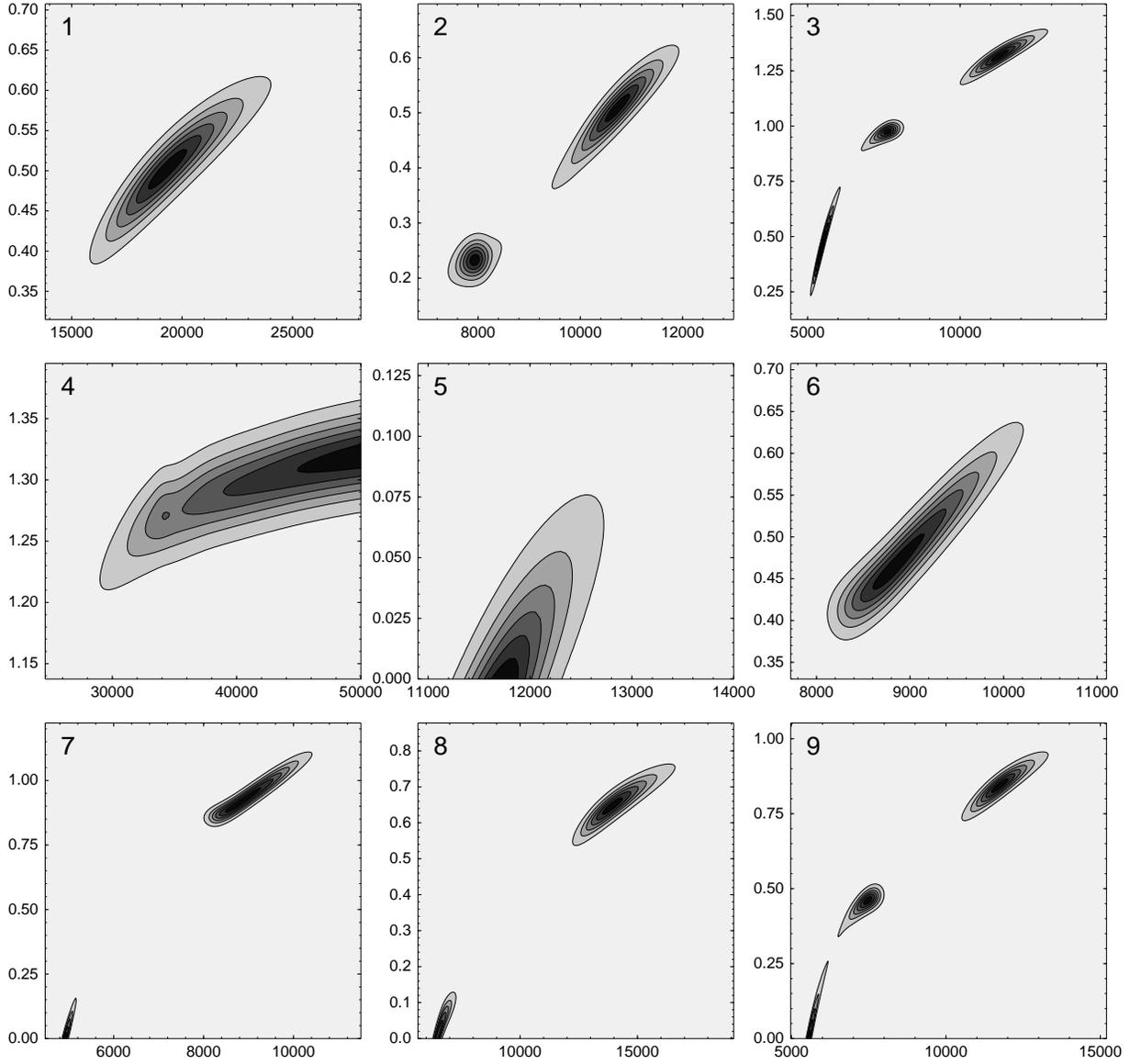}}
\caption{Likelihood contour plots produced by CHORIZOS for the nine examples
described in the text and shown in Fig.~\ref{ubvcolorcolor2}. The $x$ axis
corresponds to the temperature in K and the $y$ axis to \ebv.}
\label{ubv1}
\end{figure}

	Given the complexity and non-linearity of the general problem,
different topologies can exist, leading to different solution types as a
function of the measured magnitudes. In order to visualize them, we analyze
the well-known example of using Johnson $U-B$ and $B-V$ colors to measure 
stellar temperature and reddening ($M=N=2$). 
We show in the left panel of Fig.~\ref{ubvcolorcolor1}
the locations in 2-D color space of unreddened main-sequence stars using as 
model atmospheres those of \citet{Kuru04} with $Z=0.0$. The
Kurucz atmospheres were extinguished using a \citet{Cardetal89} law with
$R_{5495} = 3.1$ from $\ebv = 0.0$ to 5.0\footnote{\ebv\ and
$R_{5495}$ are the monochromatic equivalents to $E(B-V)$ and $R_V$, respectively.
4405 and 5495 are the assumed central wavelengths (in \AA) of the $B$ and $V$ 
filters, respectively. Monochromatic quantities are used because $E(B-V)$ and 
$R_V$ depend not only on the amount and type of dust but also on the stellar 
atmospheres.}; five different values of the temperature are shown. We also show 
in Fig.~\ref{ubvcolorcolor2} a similar plot but with more temperature values 
(here some of the symbols have been omitted for clarity). Nine examples of 
measured magnitudes are shown in Fig.~\ref{ubvcolorcolor2}, each one of them 
with different measured magnitudes $U$, $B$, $V$, but with the same uncertainty
$\sigma_U = \sigma_B = \sigma_V = 0.026$ in each case. The corresponding
solutions (as calculated by CHORIZOS) are shown in Fig.~\ref{ubv1} as 
likelihood contour plots. The examples have been selected to reflect the 
different solution types.

\begin{itemize}
  \item Example 1 is the ideal observational situation: the measured colors 
	correspond to a unique solution and the inclusion of uncertainties leads
	to a single set of connected solutions around it. 
	We can see that because the corresponding
	shaded area in Fig.~\ref{ubvcolorcolor2} is crossed only by blue lines,
	which correspond to models with $T=9\,250-50\,000$ K and this leads to
	a single connected region in Fig.~\ref{ubv1}. 
  \item Example 2 falls in a region where, due to the change in direction
	experienced by the zero-extinction color-color curve around 
	$T=9\,000$ K, two different sets of connected solutions exist. Its
	shaded area in Fig.~\ref{ubvcolorcolor2} is crossed by both blue and 
	green lines and tracing them back to the zero-extinction case we arrive 
	at two possible different ranges of values for the temperature. 
  \item Example 3 falls in the region in Fig.~\ref{ubvcolorcolor2} to the upper 
	right from where the zero-extinction color-color curve has changed 
	direction for the second time. As a result, the region is crossed by 
	lines corresponding to three temperature ranges (represented in 
	Fig.~\ref{ubvcolorcolor2} in blue, green and red, respectively) and three
	different sets of connected solutions are possible. This translates into
	three peaks in Fig.~\ref{ubv1}.
  \item Examples 4, 5, and 6 correspond to the situation where the measured $U-B$
	and $B-V$ colors are incompatible with any of the models but the inclusion
	of the corresponding uncertainties generates a single connected region 
	in Fig.~\ref{ubv1}. In the case of examples 4 and 5, the nearest edge
	of the $N-$volume of allowed colors in Fig.~\ref{ubvcolorcolor2} 
	corresponds to an extreme in $N-$dimensional parameter space (maximum
	temperature - 50\,000 K - for example 4, minimum \ebv\ - 0.0 -
	for example 5). Therefore, the corresponding likelihood contour plots in 
	Fig.~\ref{ubv1} show abrupt edges. On the other hand, the nearest edge
	of the $N-$volume of allowed colors for example 6 does not correspond to
	such an extreme; instead, it is caused by the change in direction of the
	zero-extinction color-color curve around $T=9\,000$ K. Therefore, its
	likelihood contour plot does not show an abrupt edge.
  \item For example 7 the measured $U-B$ and $B-V$ values fall outside the 
	$M-$volume of allowed colors but in this case there are two nearby 
	boundaries, one of the same type as that of example 5 and another one of the
	same type as that of example 6. This translates into two peaks in Fig.~\ref{ubv1},
	one with an abrupt edge and another one without it.
  \item Examples 8 and 9 correspond to the cases where at least one solution exists
	for the measured colors (one for example 8, two for example 9). Here, the 
	inclusion of uncertainties not only generates a connected region around 
	each one of them but also a new one, leading to the existence of two peaks 
	in Fig.~\ref{ubv1} for example 8 and of three peaks for example 9.
  \item Finally, though we have not plotted them, we can describe two possible 
	additional situations. One would be the case where the shaded region in 
	Fig.~\ref{ubvcolorcolor2} was outside the $M-$volume of allowed colors
	and far from an edge. That would indicate that the data and the models 
	are incompatible. Another situation would be reached by increasing the
	uncertainties in e.g. example 2. Then, we could have two possible solutions
	for the measured values of $U-B$ and $B-V$ but only a single set of connected
	solutions, since the shaded region in Fig.~\ref{ubvcolorcolor2} would extend
	to the line defined by the turnaround point in temperature around 9\,000 K.
\end{itemize}

	We can group our examples as a function of their solution into the three 
categories described in the introduction: unique, multiple, and non-existent. It is important
to differentiate between cases with unique and multiple solutions for the following reason. 
For unique solutions (e.g. example 1), the derived parameters can be correctly characterized 
by a single set of mean values, $\overline{T}$ and $\overline{\ebv}$, and a 
covariance matrix defined by $\sigma_T$, $\sigma_{E(4405-5495)}$, and 
$\sigma_{T,E(4405-5495)}$. In principle, one could do the same for multiple
solutions (e.g. example 2), but that would not be a correct characterization, since in 
those cases there are two or more peaks in the likelihood contour plot. In those cases, given
the strong deviations from a Gaussian distribution, it would be misleading to give mean 
parameter values, since those may fall in regions of very low probability while the main
peaks may be located at distances around or above 1 $\sigma$ from them (see. e.g. example 2, 
where the mean values are $T=9\ 910$ K, $\ebv = 0.383$, and the standard deviations 
$\sigma_T = 1\,350 $ K and $\sigma_{E(4405-5495)} = 0.139$.

	It is also important to note that, even in those cases where a unique solution is
found, there is typically a strong correlation between the parameters. This is seen in 
Fig.~\ref{ubv1} in that, if we were to approximate each peak by an ellipsoid, the two
principal axes would be inclined with respect to the $x$ and $y$ axes. This effect is 
commonly referred to in the literature as a degeneracy between the two parameters. E.g.,
for example 6 we may say that $T$ and \ebv\ are degenerate because our colors are
compatible with a single set of connected solutions where likely deviations from the mean 
require either (a) an increase in both temperature and reddening or (b) a decrease in both, 
but not e.g. an increase in temperature and a decrease in reddening.

\subsection{$M>N$ and solution existence}

\begin{figure}
\centerline{\includegraphics*[width=0.65\linewidth]{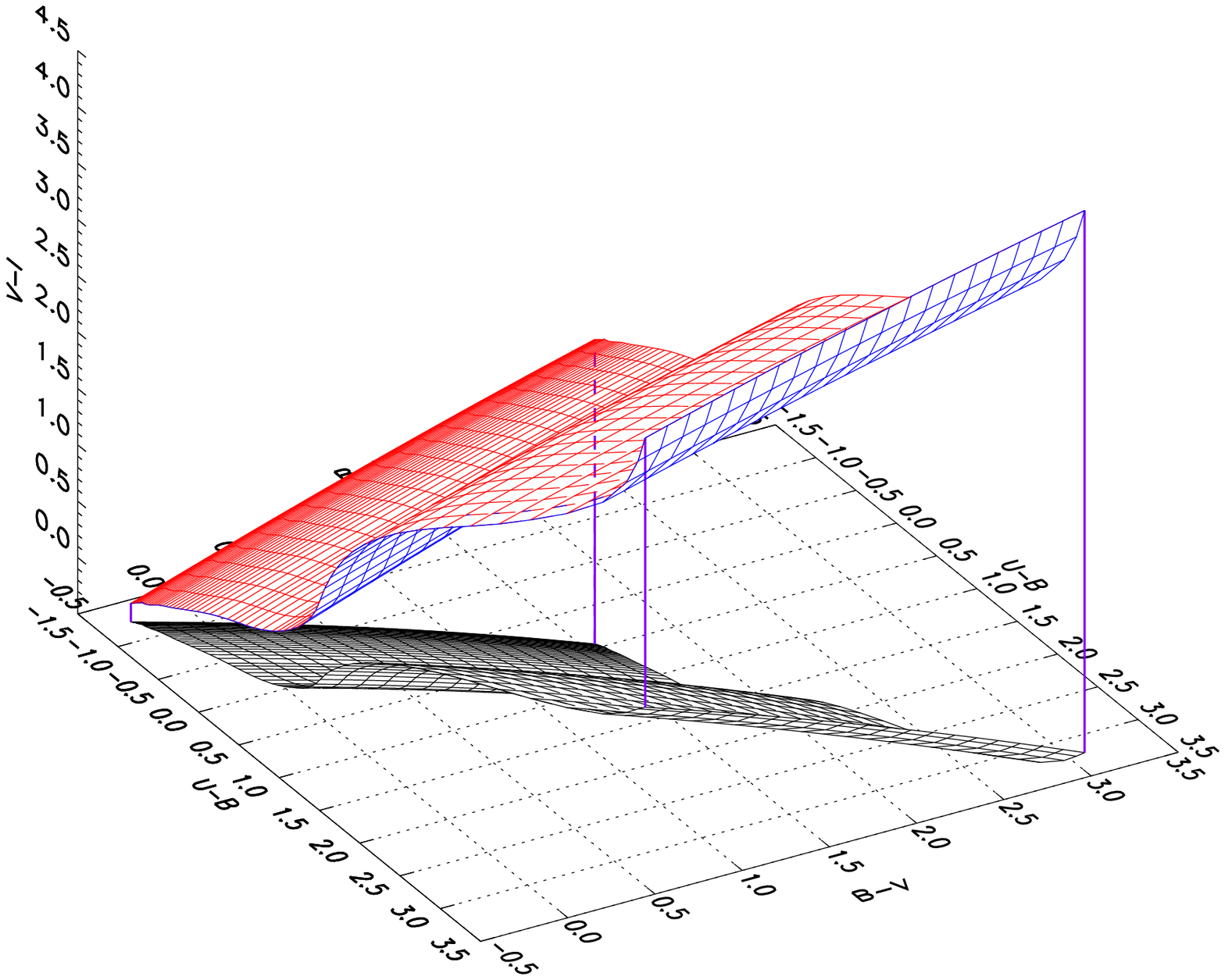}
          \ \includegraphics*[width=0.31\linewidth,bb=28 -200 566 566]{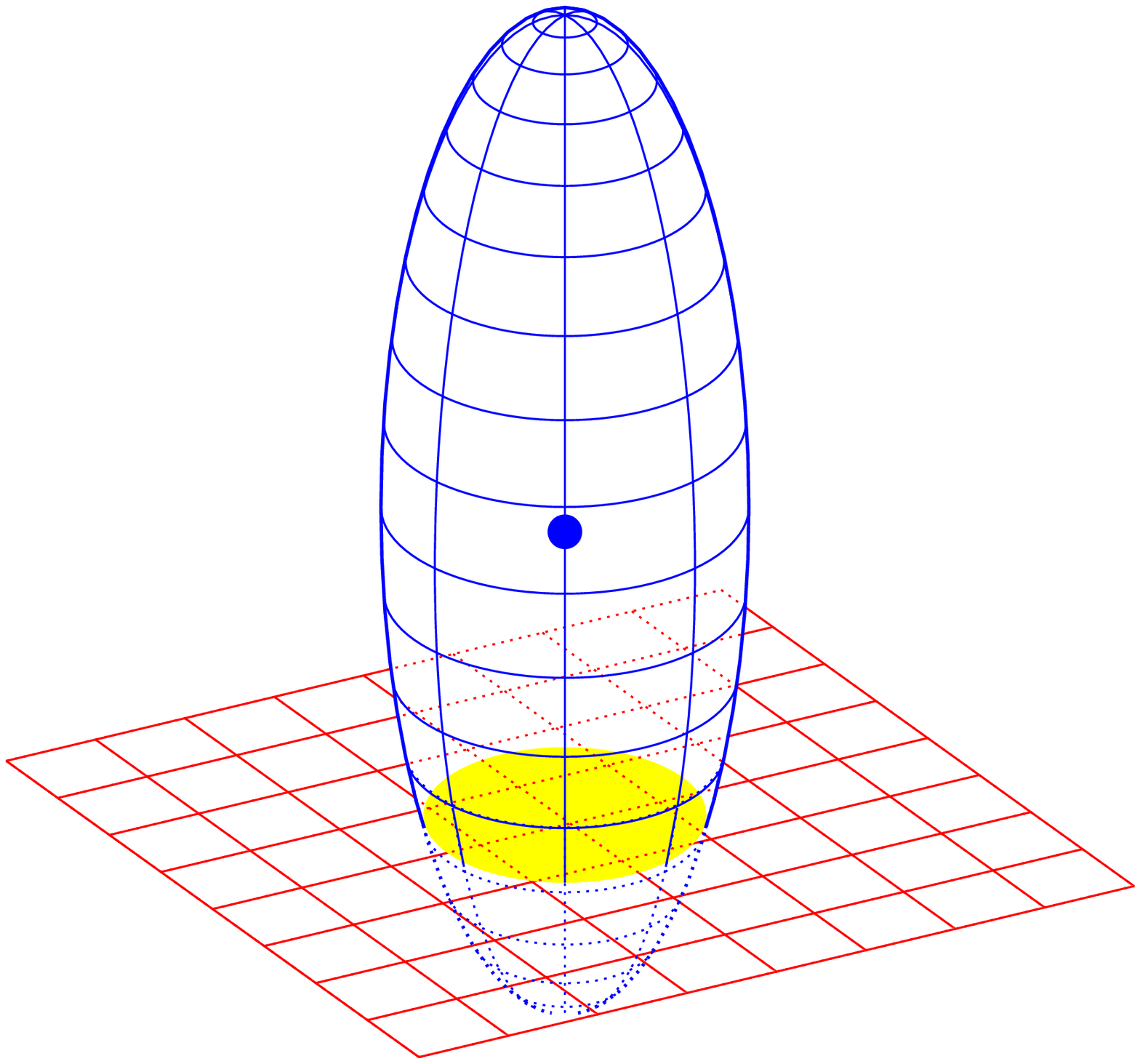}}
\caption{(left) $V-I$ vs. $B-V$ vs. $U-B$ 3-color plot for $Z=0.0$, main-sequence
Kurucz atmospheric models. The range plotted covers  $T=3\,500-50\,000$ and $\ebv=0.0-2.0$
and the extinction law used is that of \citet{Cardetal89} with $R_{5495}=3.1$.
The color surface marks the location in 3-color space while the black one is the projection
onto the $(U-B)$-$(B-V)$ plane (see Figs.~\ref{ubvcolorcolor1}~and~\ref{ubvcolorcolor2}).
(right) Basic topology for an $M=3$, $N=2$ case such as the one on the left panel. Given that 
$M>N$, the measured colors
(blue circle) always lies outside the solution surface (red grid). However, including the
uncertainty ellipsoid (blue grid), yields an intersection surface (solid yellow) of likely 
solutions.}
\label{fig3d}
\end{figure}

	The introduction of additional colors and parameters beyond two is straightforward
if we keep $M=N$ (though not as easy to plot). Instead of dealing with the intersection between
ellipses and regions of a plane, we have to find the intersection between $M-$ellipsoids and
regions of an $N-$ (or $M-$) dimensional space, which does not introduce any new behavior from the
topological point of view in our description. The situation is different if $M>N$. Take as an
example $M=3$ and $N=2$. There, we have that the two available parameters generate a surface 
($N-$volume) of possible solutions inside the volume ($M-$volume) of all possible color
combinations. Given the difference in dimensions, a given set of three measured colors will
always fall outside the solution surface. However, adding uncertainties to the measured colors
will generate a finite ellipsoid ($M-$ellipsoid) in color space that, barring problems with 
the data or the models, should intercept the solution surface. This is represented on the 
right panel of Fig.~\ref{fig3d} for the simple case where the solution surface is a plane 
perpendicular to the major axis of the uncertainty ellipsoid. For more complex solution 
surfaces, such as the one in the left panel of Fig.~\ref{fig3d}, the intersections can 
have more complex shapes but the topological classification into unique (one connected
intersection surface), multiple (two or more intersection surfaces), and non-existent (no
intersection) solution sets remains unchanged. 

	Therefore, for $M>N$ we cannot have strict solutions but only approximate ones. We can
characterize those by defining $\chi^2$ for the case of uncorrelated uncertainties as:

\begin{equation}
\chi^2 = \sum_{m=1}^{M}\frac{\left(c_m-c_{m,\rm mod}\right)^2}{\sigma_m^2}, \label{chi1}
\end{equation}

\noindent where $c_{m,\rm mod}$ are the model colors and $\sigma_m$ the measured color
uncertainties. For Gaussian uncertainties, the likelihood is then given by 
${\cal L} = \exp(-\chi^2/2)$ and maximizing it is equivalent to finding the model(s) that
minimize(s) $\chi^2$. It can be shown (see e.g. \citealt{Goul03}) that the expected 
value of \chimin\ is $M-N$, with a standard deviation of $\sqrt{2(M-N)}$. Note that those
are the results expected even for the case $M=N$, where we should find a model with the same
exact colors as those measured and, therefore, have a \chimin\ of exactly zero. Our
definition of solution existence for $M>N$ would be to have a \chimin\ reasonably close
to $M-N$ in units of $\sqrt{2(M-N)}$.

	An exception to the last statement should be noted. If the measured value lies close
to an $(N-1)-$dimensional edge of the solution $N-$volume, then the value of \chimin\ could 
be larger than $M-N$. An easy way to see this is with the most extreme case of $M=N$. In 
examples 4 to 7 of the previous section we saw that it is reasonable to have measured colors
incompatible with the model ones if they fall just outside the edge of the region spanned
by the models. In those situations where the data is close to such an edge, it is then 
normal to find somewhat higher values of \chimin: for the specific case $M=N$, one should 
then only reject the existence of a solution if $\chimin \gg 1$.

	It is still possible to find data for which there are no solutions, either for the
$M=N$ case or for the $M>N$ ones. The following is a checklist of possible causes, some
general and some specific to astronomical photometry:

\begin{itemize}
  \item Model validity: the models may be an incorrect description of the real SED.
  \item Model applicability: the type of object or the parameter range selected may be 
	incorrect.
  \item Model approximations: make sure that you are deriving your colors the right way.
	Common errors are using the wrong extinction law and neglecting curvature 
	effects in extinction trajectories for large values of \ebv\ 
	(see Fig.~\ref{ubvcolorcolor1}).
  \item Photometric quality: were the data properly acquired and calibrated?
  \item Zero-point calibration: some filters require small magnitude corrections before
	their magnitudes can be used for spectrophotometry (see e.g. \citealt{Coheetal03}).
  \item Filter-throughput errors: make sure that the throughputs of your filter are
	correctly characterized. Be on the alert for filters with red/blue leaks and long 
	tails.
  \item Color transformations: beware of transforming from one filter system to another.
	Whenever possible, use the throughput definitions for the filters with which the
	data was acquired.
\end{itemize}

\section{Techniques}

	Apparently, two different techniques should be needed to solve the two different
cases defined in the previous section. In practice, both can be solved using a $\chi^2$
minimization algorithm. That is so because for $M=N$, the algebraic solution to the system of
equations is also the solution that makes $\chi^2 = 0$, which is obviously the minimum
possible value for $\chi^2$. Furthermore, since ${\cal L} = \exp(-\chi^2/2)$ independently of the
number of degrees of freedom, we can use that definition to estimate the uncertainties
in our parameters not only for $M>N$ but also for $M=N$.

	CHORIZOS works by computing $\chi^2$ over the $N-$dimensional parameter grid and calculating 
the corresponding likelihood at each grid point. The current version of the code is written in IDL and
handles up to $N=4$: two parameters from the intrinsic properties of the SED family plus two 
extinction-related parameters, \ebv\ and extinction law
type. The latter includes the $R_{5495}$-dependent family of extinction laws of \citet{Cardetal89}, 
the average LMC and LMC2 laws of \citet{Missetal99}, and the SMC law of \citet{GordClay98}. CHORIZOS
starts by reading the unreddened SED models, extinguishing them, and obtaining the synthetic
photometry at each point in a (coarse) 4-D grid. 
This is done in order to correctly deal with non-linear extinction effects. This preliminary step 
needs to be executed only once and the result can be stored in the form of binary FITS tables 
for later use. Currently, CHORIZOS includes pre-calculated tables for Kurucz \citep{Kuru04},
Lejeune \citep{Lejeetal97}, and TLUSTY \citep{LanzHube03} 
stellar models and Starburst 99 \citep{Leitetal99} cluster
models using a total of 78 filter passbands (including the Johnson, Str\"omgren, 2MASS, and several
HST instrument systems). The two intrinsic parameters are temperature and gravity for the stellar
models and age and metallicity for the cluster models. More models and passbands will be included in 
the future and the user will also be able to add his/her own. 

	The program first reads the photometry from a user-provided table, as well as a series of
input parameters, such as which model family should be used and how fine a parameter grid should be 
employed. At this point, the user can restrict any of the four parameters for any of the objects 
(stars, clusters or galaxies) in the input table, using either a specific value or a range between a 
minimum and a maximum. CHORIZOS then 
reads the model photometry from the previously calculated tables and interpolates to the user-selected
(fine) grid. For each object, the likelihood is calculated at each grid point, the average values
of the four parameters and their (output) 4$\times$4 covariance matrix are computed, and the result is
written in an individual file for each star. This file is read by a second program which 
produces the graphical output and allows for further manipulation of the results, such as calculation of
stellar absolute magnitudes or cluster stellar masses.

	There are several issues regarding algorithm details and results interpretation for a program
of this type that need to be discussed. First, if one measures magnitudes directly and then calculates
two different colors that include one filter in common (e.g. $B$ in the example discussed in the
previous section), then the probability distributions for the two colors will be correlated (see,
e.g. the data plotted in Fig.~\ref{ubvcolorcolor2}). It can
be easily shown that for the case of the two colors $X-Y$ and $Y-Z$ formed from the three filters 
$X$, $Y$, and $Z$, the input (or color) covariance matrix is:

\begin{equation}
{\cal C}_{X-Y,Y-Z} = {\rm cov}(X-Y,Y-Z) = \left(\begin{array}{cc} 
\sigma_X^2+\sigma_Y^2 & -\sigma_Y^2           \\
-\sigma_Y^2           & \sigma_Y^2+\sigma_Z^2
\end{array}\right),
\label{cov}
\end{equation}

\noindent where $\sigma_X$, $\sigma_Y$, and $\sigma_Z$ are the uncertainties which correspond to each
filter. For correlated uncertainties, Eqn.~\ref{chi1} is no longer valid and one has to use (see e.g.
\citealt{Goul03}):

\begin{equation}
\chi^2 = \sum_{l=1}^{M}\sum_{m=1}^{M}\left(c_l-c_{l,\rm mod}\right){\cal B}_{lm}
                                     \left(c_m-c_{m,\rm mod}\right), \label{chi2}
\end{equation}

\noindent where ${\cal B}\equiv {\cal C}^{-1}$, the inverse of the $M\times M$ input (or color) 
covariance matrix. Note, however,
that in some cases two colors such as $U-B$ and $B-V$ can be considered to be in a first approximation
as uncorrelated. That would be the case if those colors are built from the combination of a large
number of independent measurements of $U-B$ and $B-V$, a situation that one may encounter when
collecting data from the literature. CHORIZOS allows either option to be used: if individual 
magnitudes are inputted, then the full input covariance matrix is utilized; if colors are chosen, then
only the corresponding diagonal terms are used. Given the relative complexity of the algorithm
implementation, 
We conducted an independent test using a Montecarlo simulation in which the measured colors were
varied according to the input covariance matrix and the solution that minimized $\chi^2$ was selected
in each case. The Montecarlo simulation was run with 1000 samples and the results were combined to
produce a likelihood map which was then compared to the CHORIZOS output using a variety of input data
and models\footnote{Note that a Montecarlo method is a valid alternative way of attacking this problem 
but it is more expensive from the computational point of view.}. Results agreed in all cases.

	Another issue is that of grid size. When calculating the solution, CHORIZOS also checks
whether the width of the resulting distribution in parameter space is comparable to that of the grid 
size near the optimum solution and warns the user if it finds that condition is not met 
so that a new run with a finer grid can be executed. 
A related issue is that of the value of \chimin\ for the case $M=N$. As we have mentioned, 
it should be zero, unless the measured colors are incompatible with the models (e.g. the solution 
falls outside the $N-$volume of allowed colors). Therefore, the value of \chimin\ can 
be used to check such a circumstance. Since we are using a grid to evaluate $\chi^2$, the minimum
value will not be zero but if the grid is fine enough it should be $\ll 1$. But what if the grid is
not fine enough? To solve this problem, after CHORIZOS finds the solution that minimizes $\chi^2$,
it interpolates the colors from the adjacent points into an ultrafine grid in order to obtain a more 
accurate value of the minimum. This value can then be used to decide whether the measured colors fall
outside the $N-$volume of allowed colors or not.

	A problem also arises when the measured colors fall exactly at the edge of the $N$-volume of
allowed colors. One example would be if in the temperature-reddening case using $U-B$ and $B-V$ the
measured colors were to correspond to those of a star with zero reddening (the resulting likelihood 
diagram could be similar to that of example 5 in Fig.~\ref{ubv1}). In that case, the resulting mean 
reddening value from the likelihood data would be greater than zero and if we were to measure a number
of stars with real \ebv\ equal to zero, we would obtain positive values in all cases, leading us
to an incorrect estimation of the value of the reddening. One solution which is implemented in
CHORIZOS as an option is to extrapolate the grid to values beyond those provided by the original
models. Using that option, the likelihood plot for example 5 in Fig.~\ref{ubv1} would show the full
ellipsoid and the correct mean value could be obtained (which in that specific case would be negative,
given the location of point 5 in Fig.~\ref{ubvcolorcolor2}). This option should be used with caution,
though, since extending the grid too much can easily lead to the inadvertent introduction of false
additional solutions. Also, in some cases the use of extrapolated values is not recommended due to the
existence of color near-degeneracies at the grid edges (e.g. the optical colors at the high-temperature 
end for O stars).

	Finally, the validity of the derived mean values and covariance matrix when the calculated
parameter distribution is far from being a Gaussian (e.g. when multiple solutions exist) has to be
analyzed. We will do this in the following section, where we discuss some sample applications of
CHORIZOS.

\section{Sample applications}

\subsection{Multiple solutions for stars using Johnson-Cousins photometry}

	As we have seen, $U-B$ and $B-V$ colors alone are not enough to provide a single solution 
under all circumstances for main-sequence stellar atmospheres using a fixed extinction law. But what 
if we use additional filters in the Johnson-Cousins set? 
If we add the $I$ filter and use $V-I$ as a third color, we 
have the situation represented on the left panel in Fig.~\ref{fig3d}. We can see that in this case
a third color eliminates most or all the multiple solutions. However, since we now have $M=3>N=2$,
the problem does not have an exact solution for an arbitrary combination of colors and we are forced
into using a $\chi^2$ minimization or a similar technique.

	We analyze here a similar case using five Johnson-Cousins filters and a total of four colors: $U-B$,
$B-V$, $V-R$, and $V-I$ to measure the temperature and reddening of a main-sequence star. 
We use $Z=0.0$, main-sequence Kurucz atmospheres, restrict the 
extinction law to be that of \citet{Cardetal89} with $R_{5495}=3.1$, and fix the uncertainties in
all five passbands to be 0.01 magnitudes. The grid was extrapolated in \ebv\ but not in $T$.
In order to compare the results obtained with the full
$UBVRI$ photometry with those of only $UBV$ photometry, We ran CHORIZOS using its test mode. In that
mode, CHORIZOS is fed the true model colors and is executed to see if it is able to reproduce them. If
there is a single well-behaved solution, the measured mean parameters should be very similar to the 
input ones; if there are multiple solutions, that may not be the case. One way to quantify the effect
is to use the mean calculated temperature and reddening, $\overline{T}$ and $\overline{\ebv}$, and their 
uncertainties estimated from their standard deviations, $\sigma_T$ and $\sigma_{E(4405-5495)}$, to 
calculate their normalized distances from the input value:

\begin{equation}    
d_T = (\overline{T}-T_{\rm input})/\sigma_T,
\label{normt}
\end{equation}

\begin{equation}    
d_{E(4405-5495)} = (\overline{\ebv}-\ebv_{\rm input})/\sigma_{E(4405-5495)}.
\label{normebv}
\end{equation}

\begin{figure}
\centerline{\includegraphics*[width=0.48\linewidth]{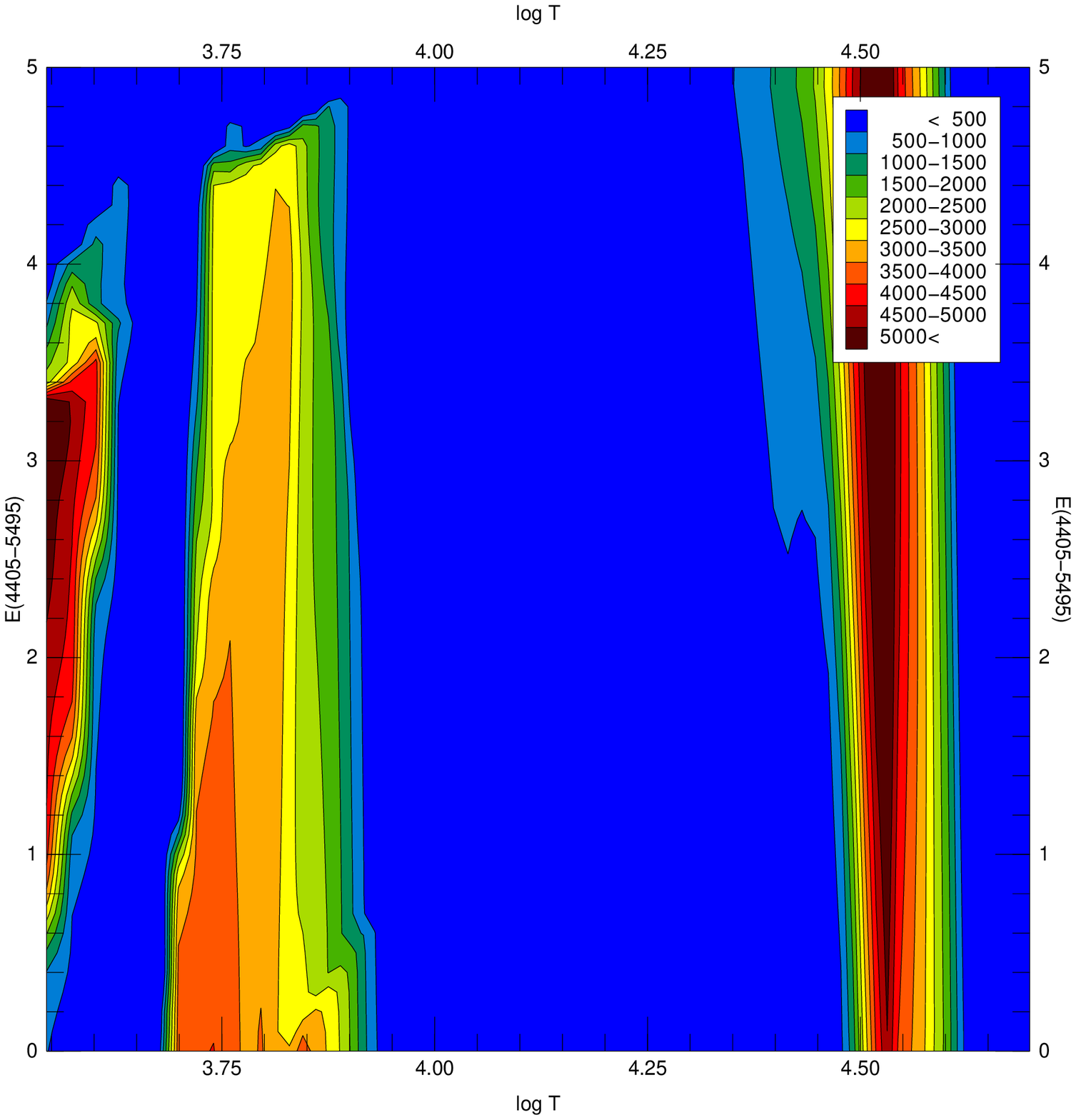}
          \ \includegraphics*[width=0.48\linewidth]{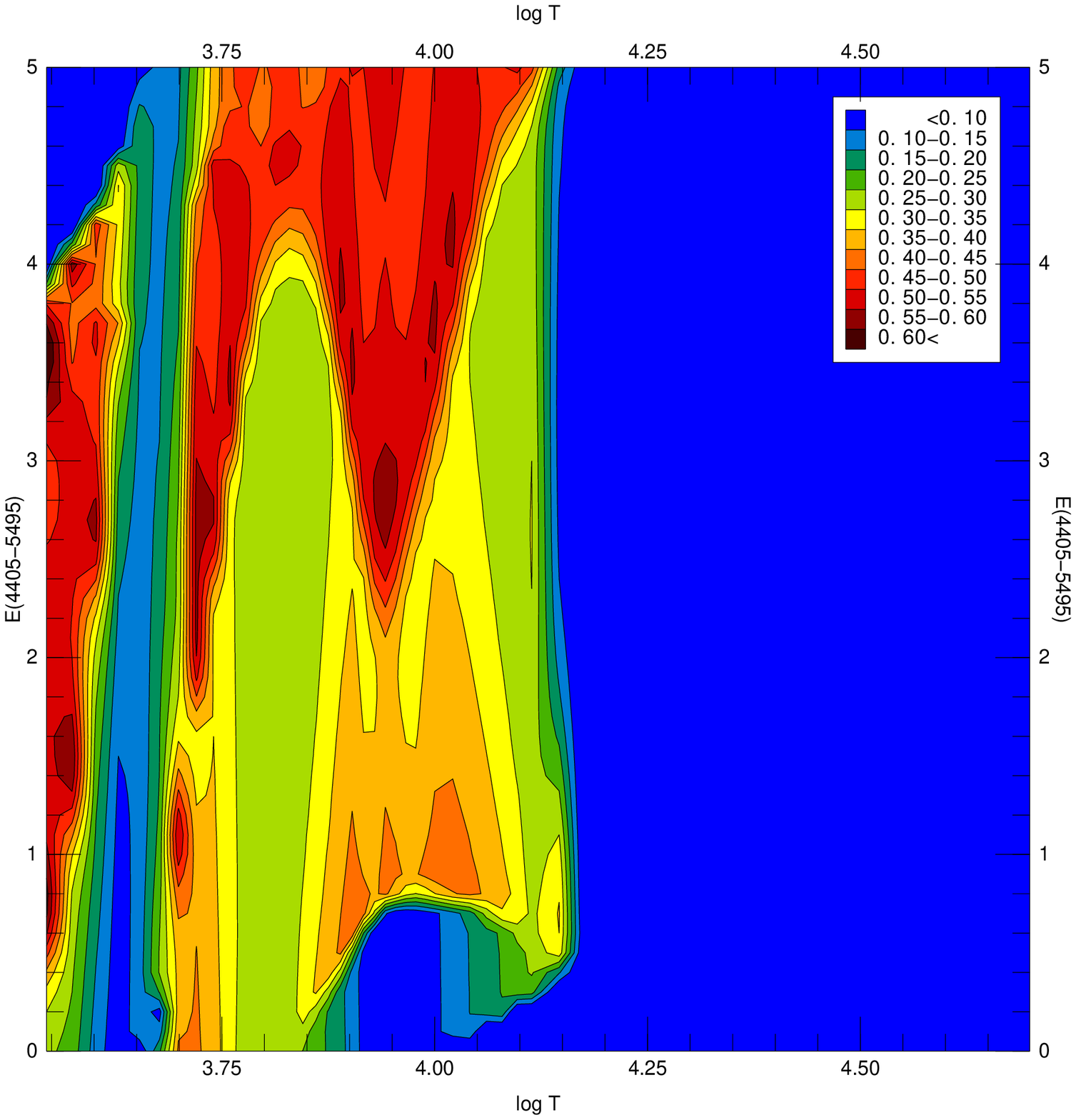}}
\centerline{\includegraphics*[width=0.48\linewidth]{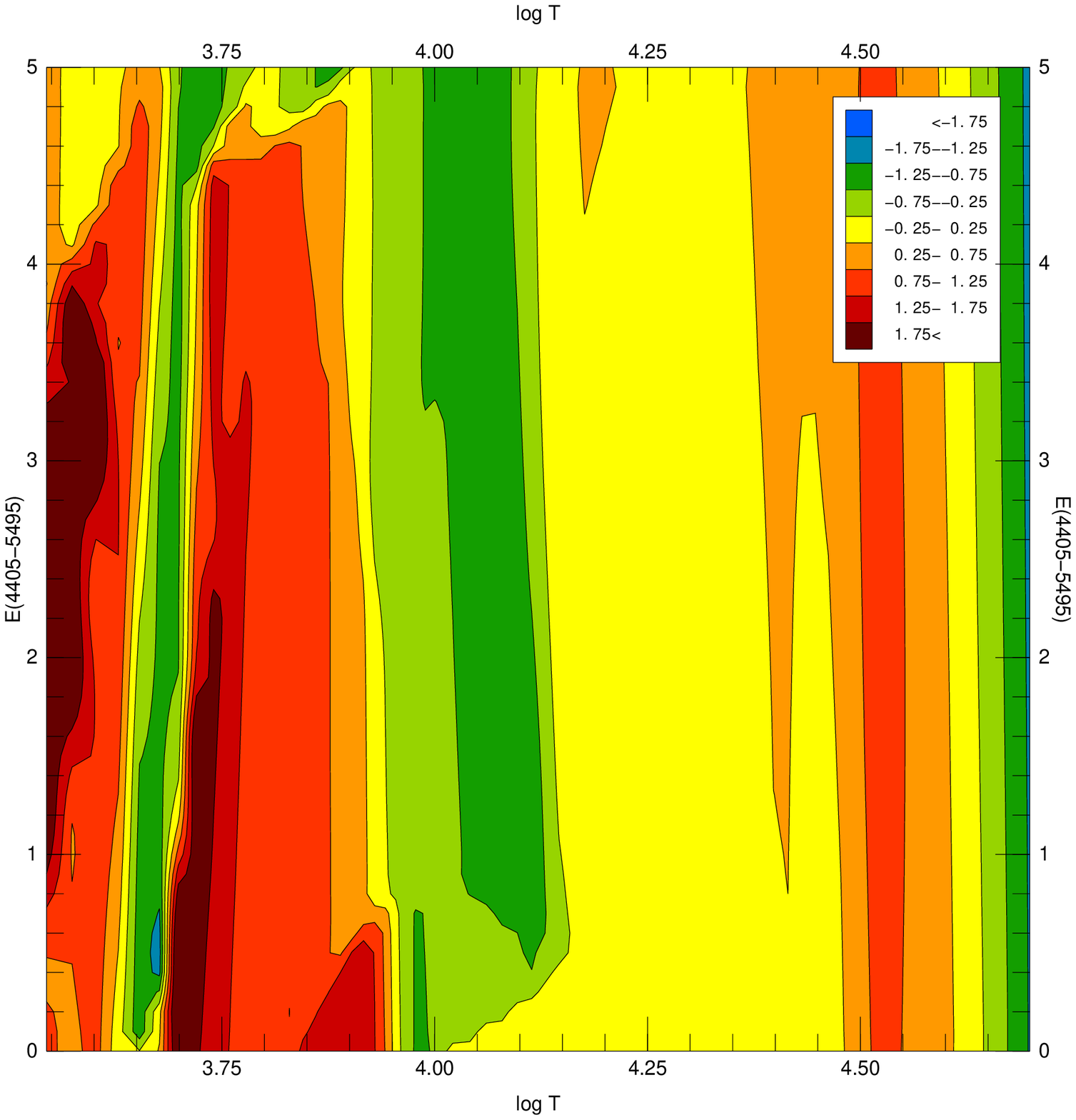}
          \ \includegraphics*[width=0.48\linewidth]{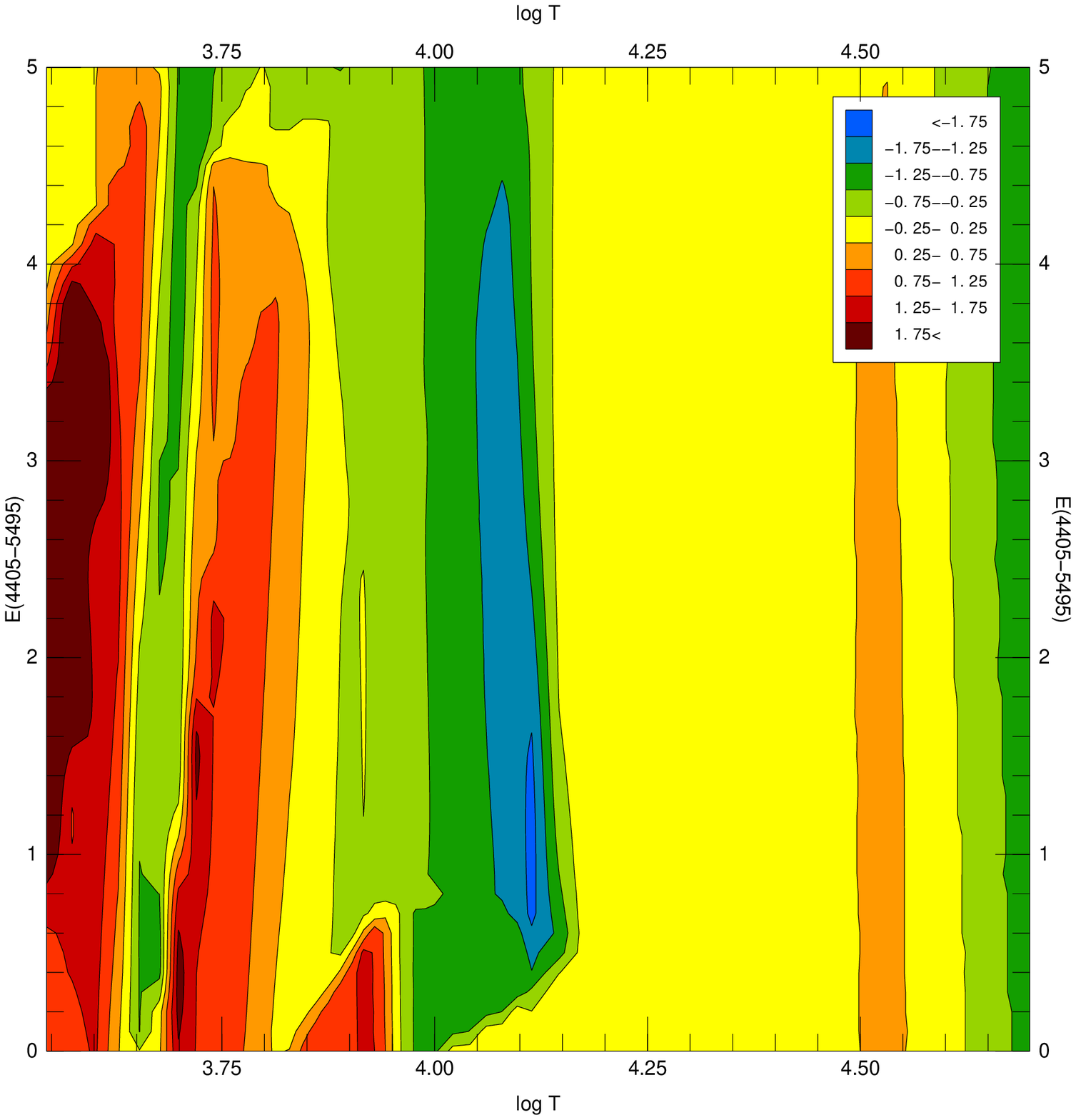}}
\caption{(top) Plots of $\sigma_T$ (left) and $\sigma_{E(4405-5495)}$ (right) 
for sample application 1 (main-sequence Kurucz models with $R_{5495} = 3.1$) as a 
function of $\log T$ (in K) and \ebv\ calculated for $UBV$ photometry 
with uncertainties of 0.01 magnitudes for each filter. (bottom) Plots of
$d_T$ (left) and $d_{E(4405-5495)}$ (right) corresponding to the same case.}
\label{sample1_ubv}
\end{figure}

	For a single well-behaved solution one expects low values of $\sigma_T$ and 
$\sigma_{E(4405-5495)}$ (unless the input uncertainties themselves are large) and, more importantly, values 
of $|d_T|$ and $|d_{E(4405-5495)}| < 1.0$ (and typically lower than 0.5). If there are two or more 
solutions, the normalized distances can easily be larger than one because for such a distribution (formed 
by e.g.  two well-separated narrow quasi-Gaussians) the maxima can be separated from the mean by more
than one standard deviation. We have plotted those four quantities for the test
case with only $UBV$ photometry in Fig.~\ref{sample1_ubv}. 

\begin{itemize}
  \item For $T> 15\, 000$ K, CHORIZOS detects the existent unique solutions, 
	as evidenced by the relatively low values of the uncertainties and
	absolute values of the normalized distances. Two minor points can be 
	noted: the somewhat larger values of $\sigma_T$ around $37\, 000$ K are
	caused by the near-degeneracy of the optical colors of O stars. Also, 
	the green region at the right border of the normalized distance plots 
	is caused by edge effects (we did not extrapolate in temperature). 
  \item For $T< 15\, 000$ K, CHORIZOS produces in general much larger values of
	the uncertainties (especially for \ebv) and the normalized distances 
	are of the order of or larger than one (in absolute value) for most
	of the region. This is caused by the existence of multiple solutions.
\end{itemize}

\begin{figure}
\centerline{\includegraphics*[width=0.48\linewidth]{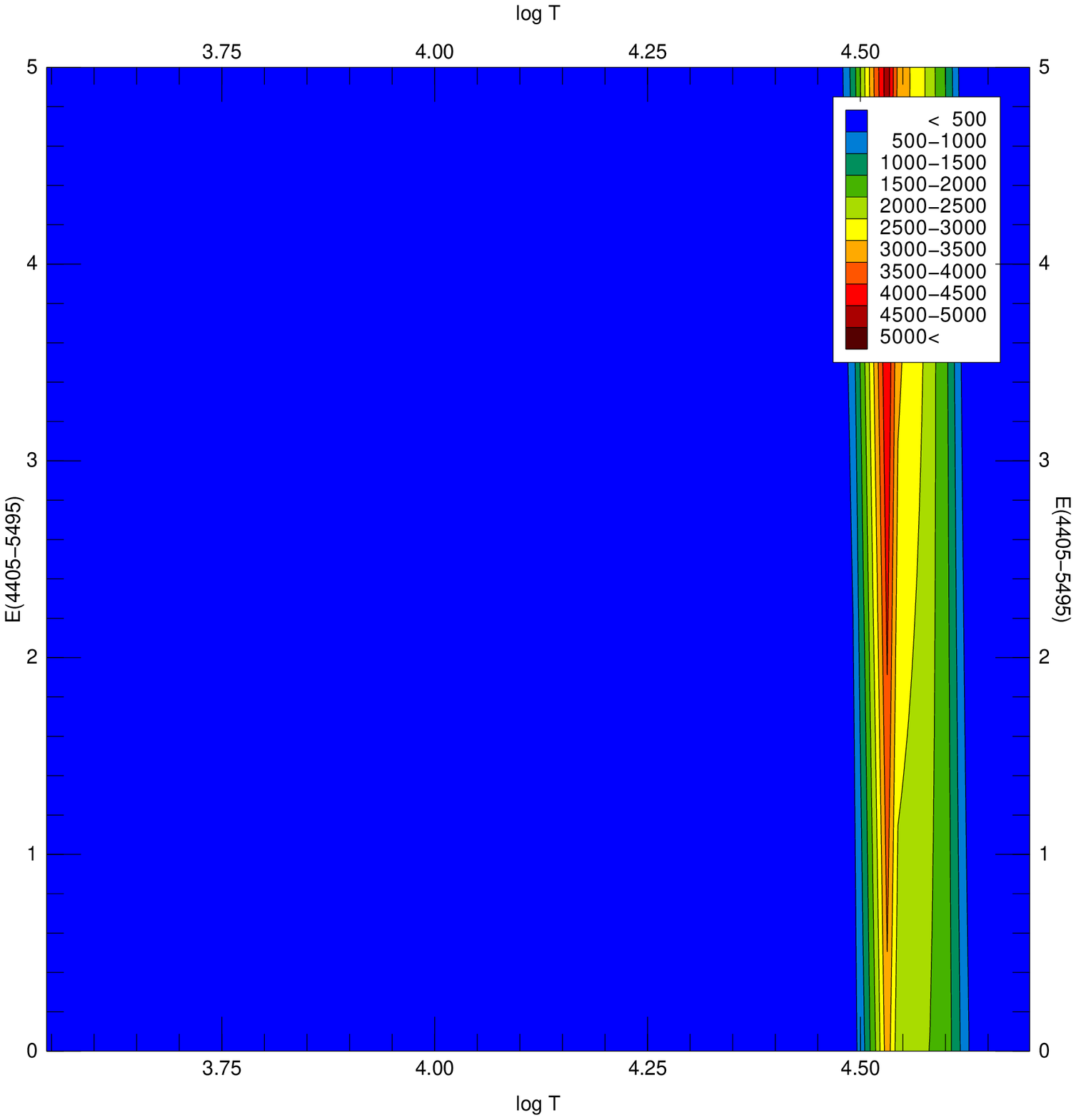}
          \ \includegraphics*[width=0.48\linewidth]{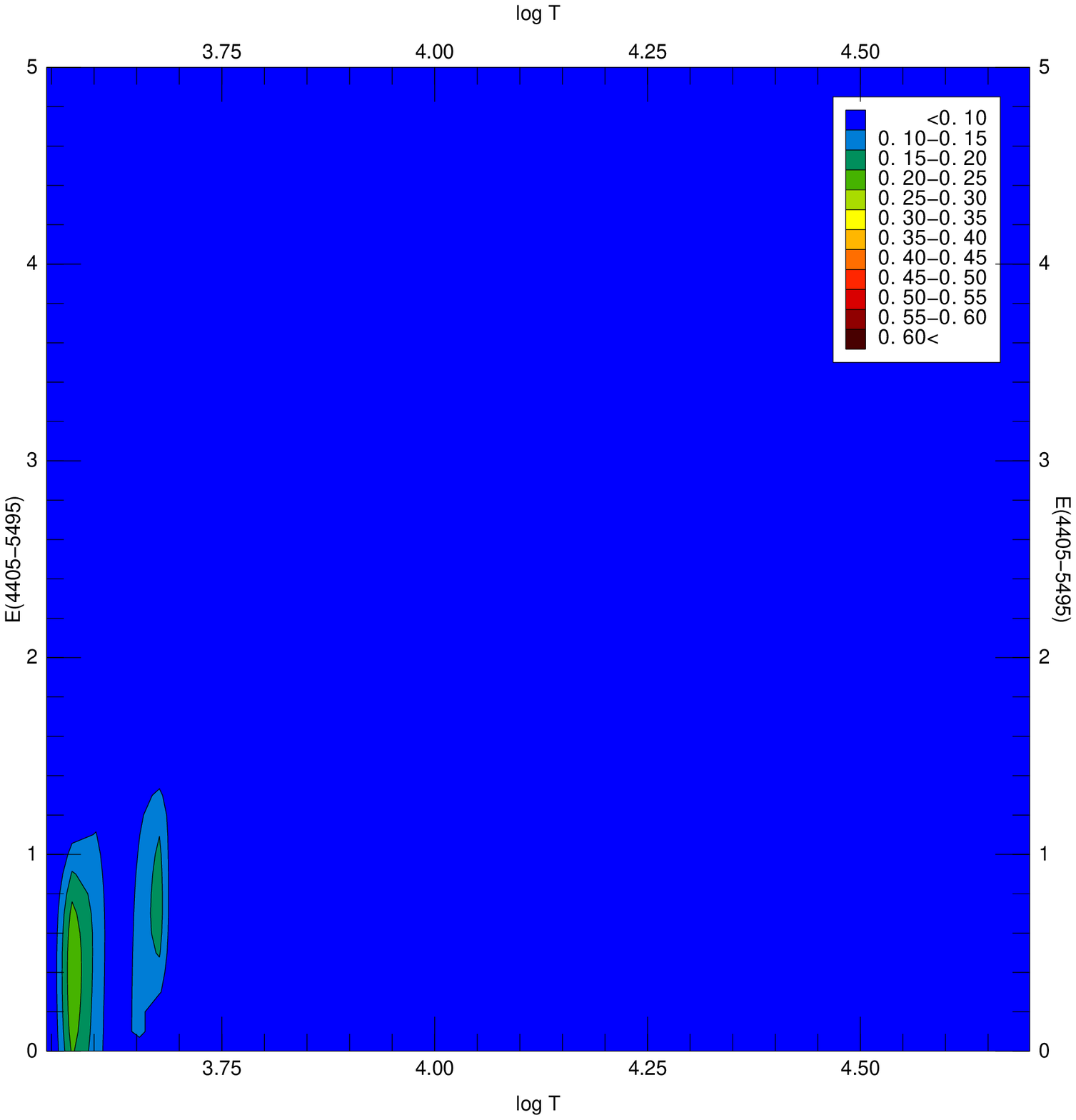}}
\centerline{\includegraphics*[width=0.48\linewidth]{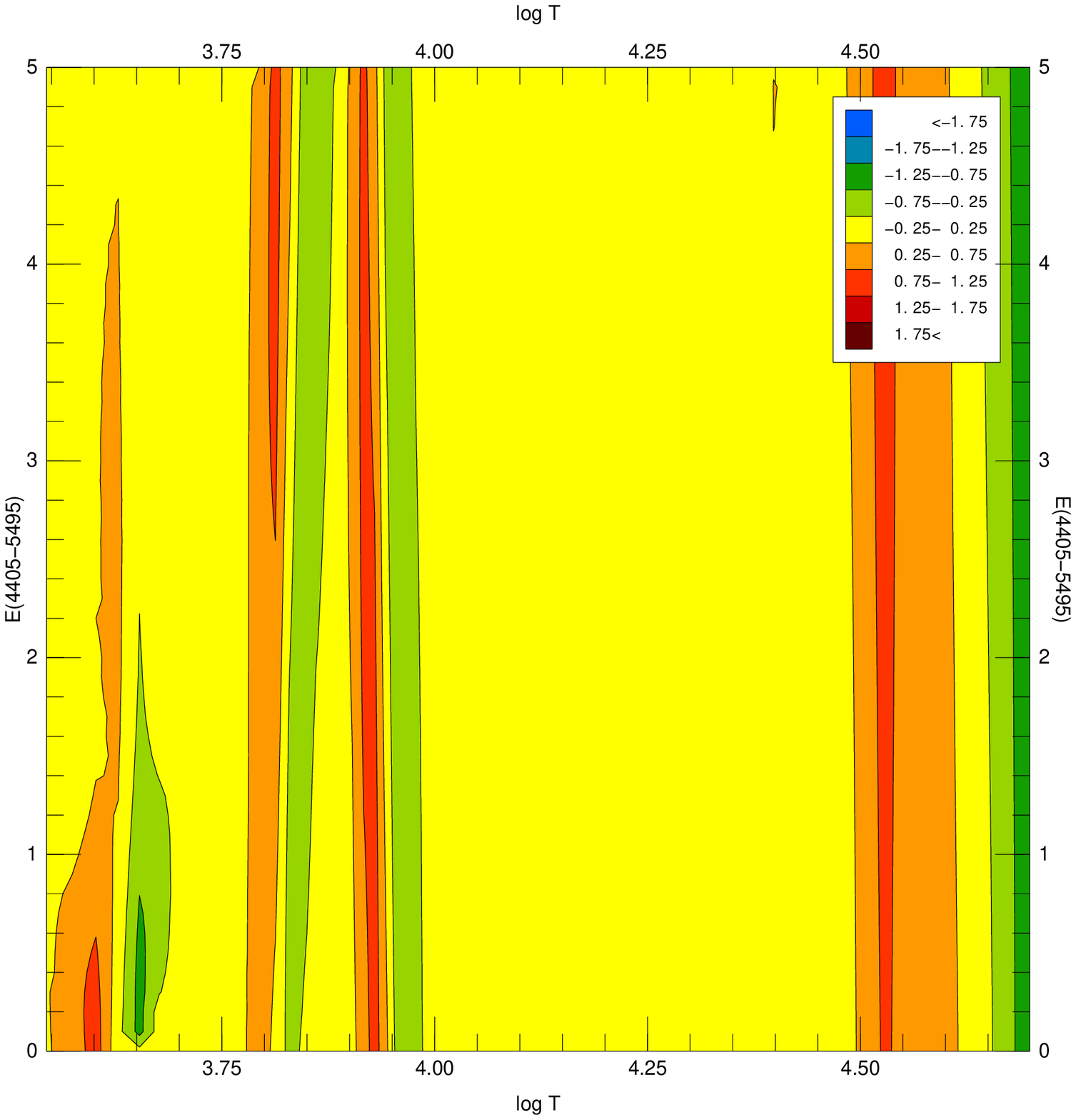}
          \ \includegraphics*[width=0.48\linewidth]{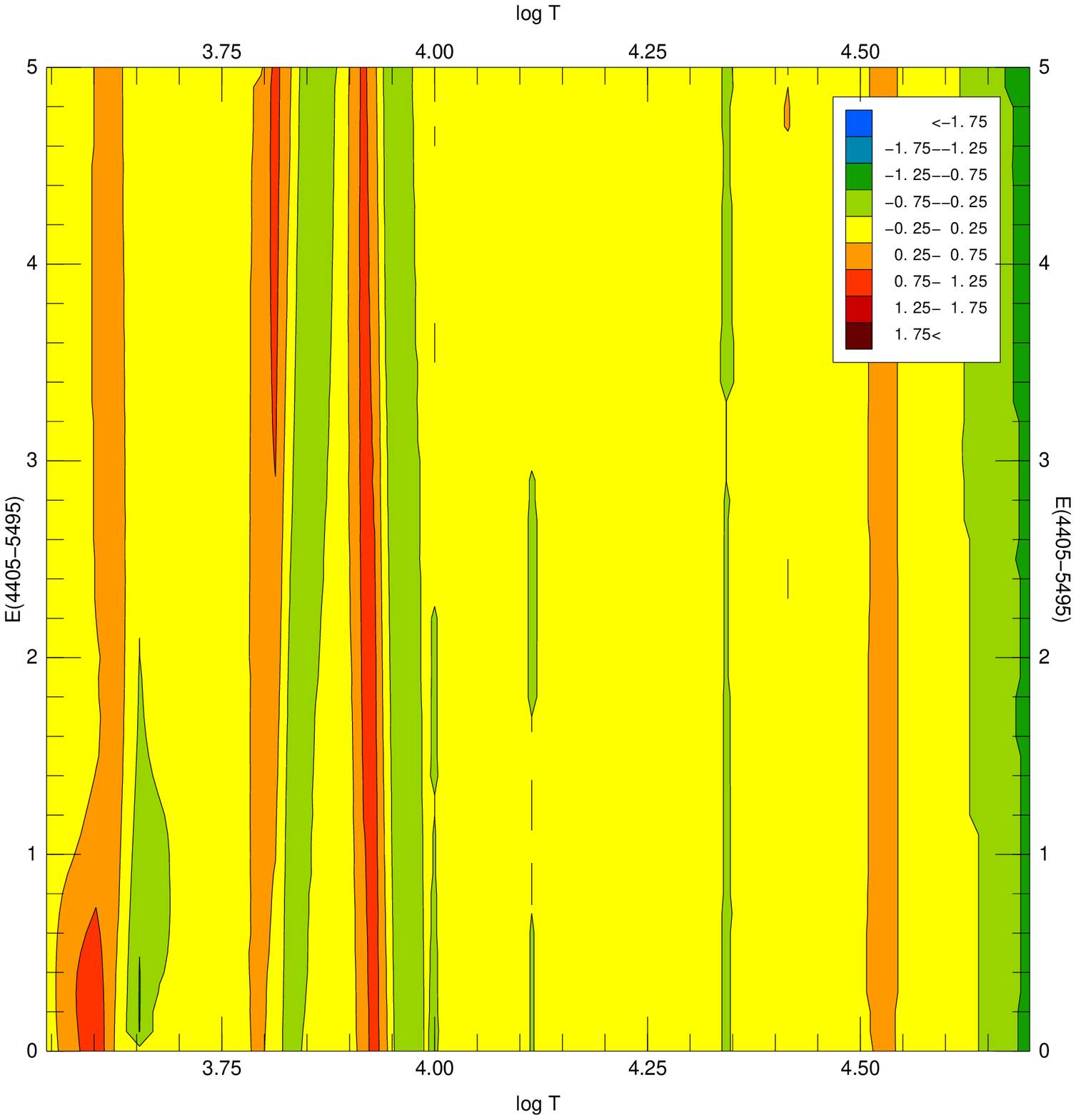}}
\caption{Same as Fig.~\ref{sample1_ubv} for $UBVRI$ photometry.}
\label{sample1_ubvri}
\end{figure}

	These results are simply what we expected. The interesting point is
that this demonstrates that when CHORIZOS is run in test mode, it can be used
to predict which parameter ranges can be measured with the existent colors and
precisions and which ones cannot be measured. We now turn to 
Fig.~\ref{sample1_ubvri}, which has the same plots but for the test case with
the full $UBVRI$ data using the same color scale.

\begin{figure}
\centerline{\includegraphics*[width=0.57\linewidth]{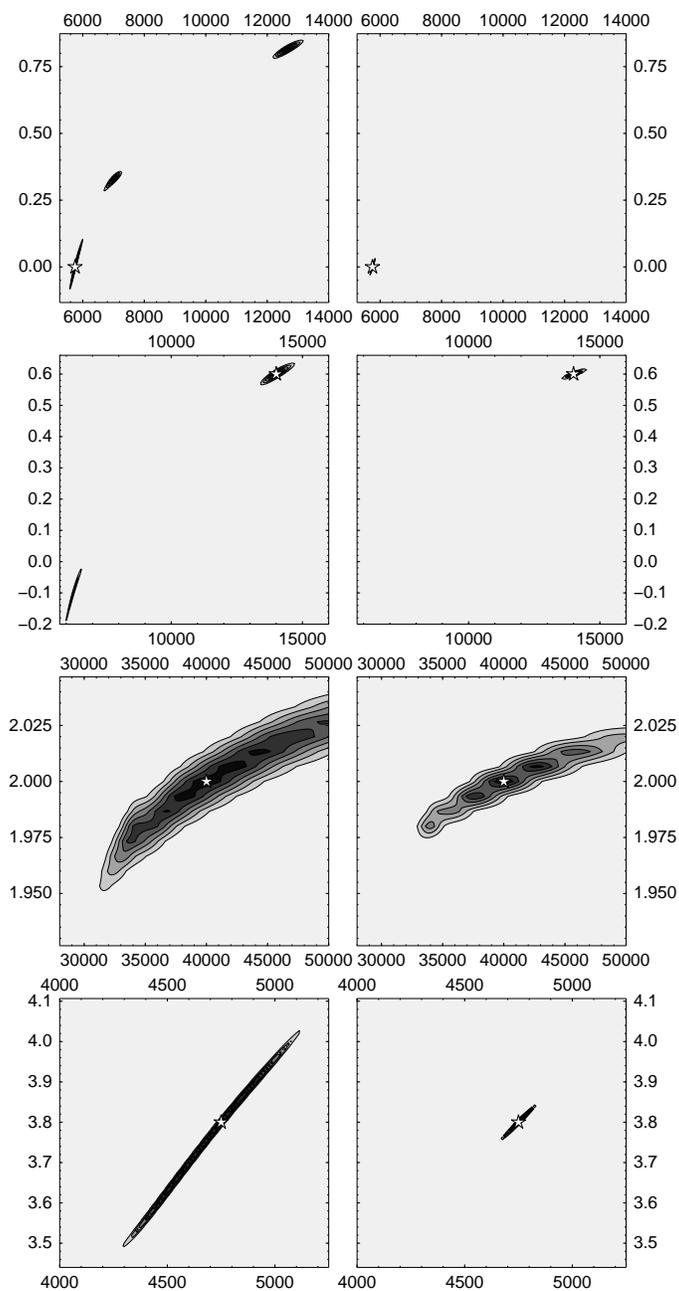}}
\caption{Likelihood contour diagrams for four selected cases in sample application 
1 using $UBV$ data (left) and $UBVRI$ data (right). Temperature (in K) is
plotted in the horizontal axis and \ebv\ in the vertical axis. A white star 
is used to indicate the true temperature and reddening. The irregularities in
the contour diagram for the third star are caused by grid-size effects.}
\label{ubv_ubvria}
\end{figure}

\begin{figure}
\centerline{\includegraphics*[width=0.60\linewidth]{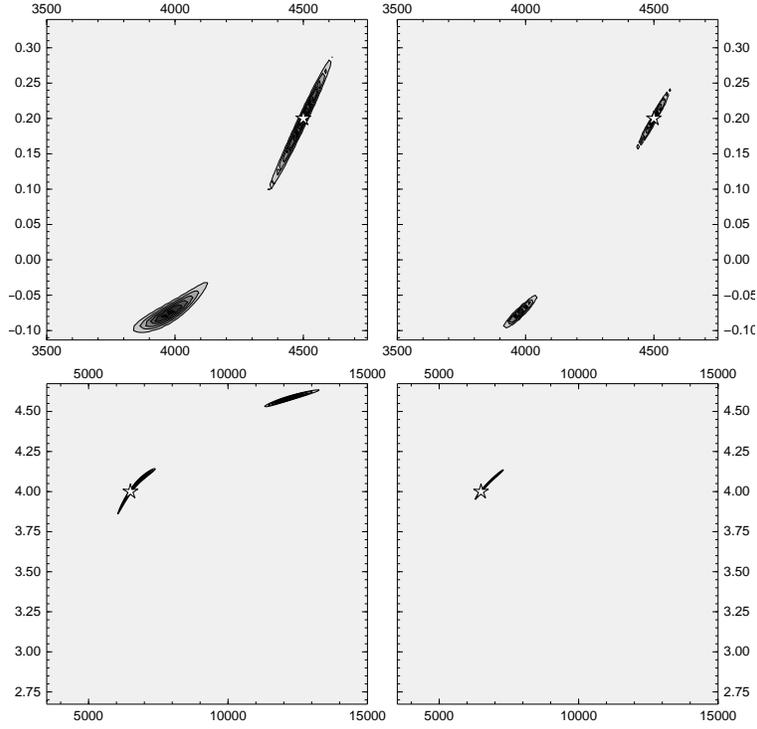}}
\caption{Same as Fig.~\ref{ubv_ubvria} for two additional cases.}
\label{ubv_ubvrib}
\end{figure}

\begin{itemize}
  \item The values of the uncertainties are much lower, being $< 500$ K and
	$<0.10$ for $\sigma_T$ and $\sigma_{E(4405-5495)}$, respectively,
	over most of the analyzed range. The only significant deviations are
	those for $\sigma_T$ around $T\approx 40\, 000$ K, where the 
	introduction of $RI$ data alleviates the color near-degeneracy but does 
	not eliminate it completely, and for $\sigma_{E(4405-5495)}$ at certain 
	regions near the lower left of the diagram; both effects are quite
	minor.
  \item The values of the normalized distances (in absolute value) are also
	much smaller, with only a few regions coming close to 1.0 (in some 
	cases due to edge effects; see, however, the comparison below with 
	Lejeune atmospheres)). 
  \item A comparison between the likelihood plots produced by CHORIZOS from
	$UBV$ and $UBVRI$ cases are shown in 
	Figs.~\ref{ubv_ubvria}~and~\ref{ubv_ubvrib}. The first of those two
	figures shows four examples where the introduction of $RI$ data
	yields excellent solutions, in some cases eliminating one of the
	multiple solutions and in others simply reducing the extent of the 
        possible range of values. The second figure shows two of the cases
	where $|d_T|$ and $|d_{E(4405-5495)}|$ are close to 1.0 even for
	the $UBVRI$ case. We can see that, even in those ``bad'' cases, the
	introduction of $RI$ data is useful in further constraining the
	values of the parameters. The reasons for the high values of the
	normalized distances are, in one case, the persistence of two
	solutions (with a narrowing of each peak) and, in the other one,
        the reduction to a single solution with an asymmetrical 
        distribution.
\end{itemize}

\begin{figure}
\centerline{\includegraphics*[width=0.48\linewidth]{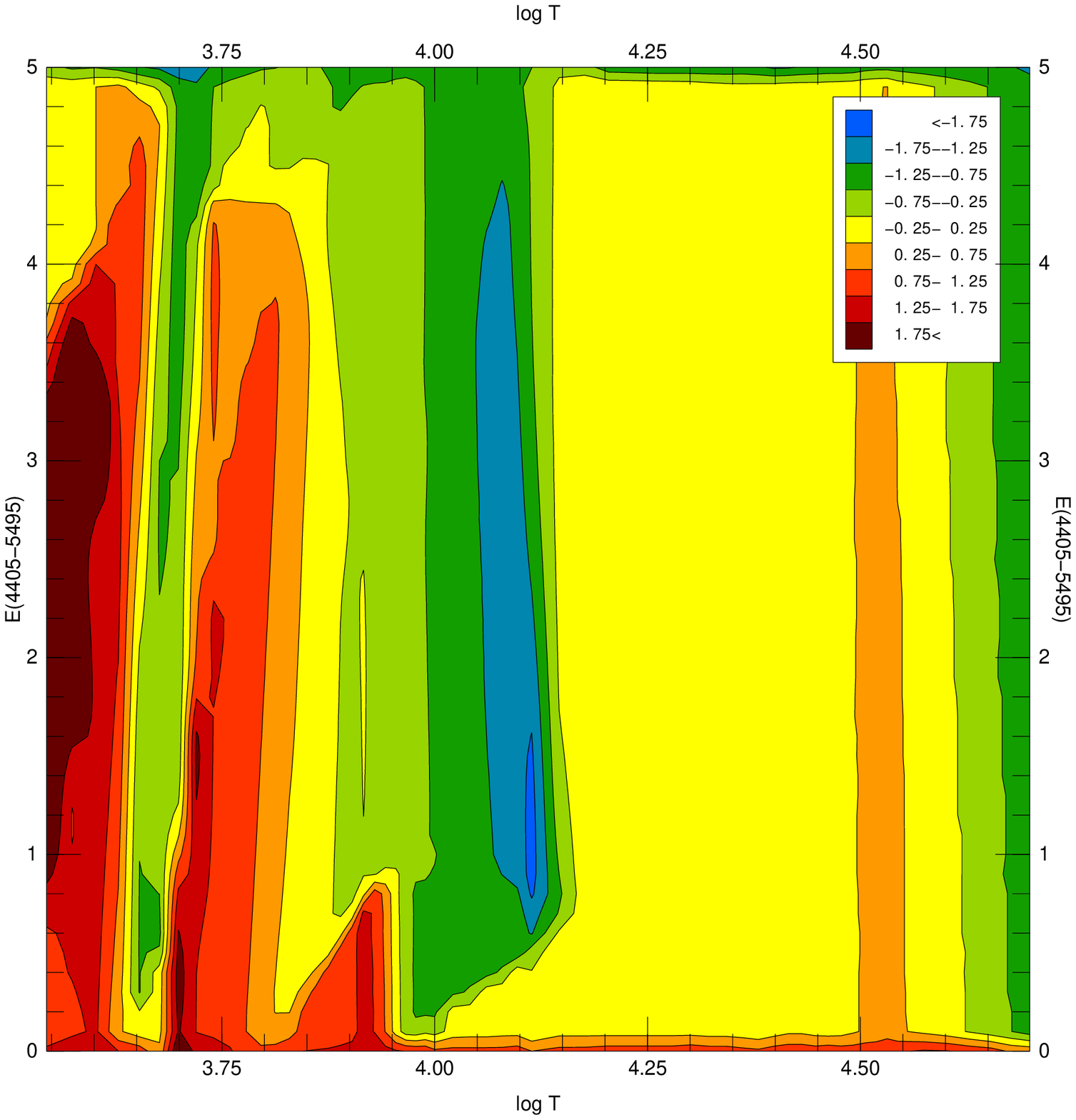}
          \ \includegraphics*[width=0.48\linewidth]{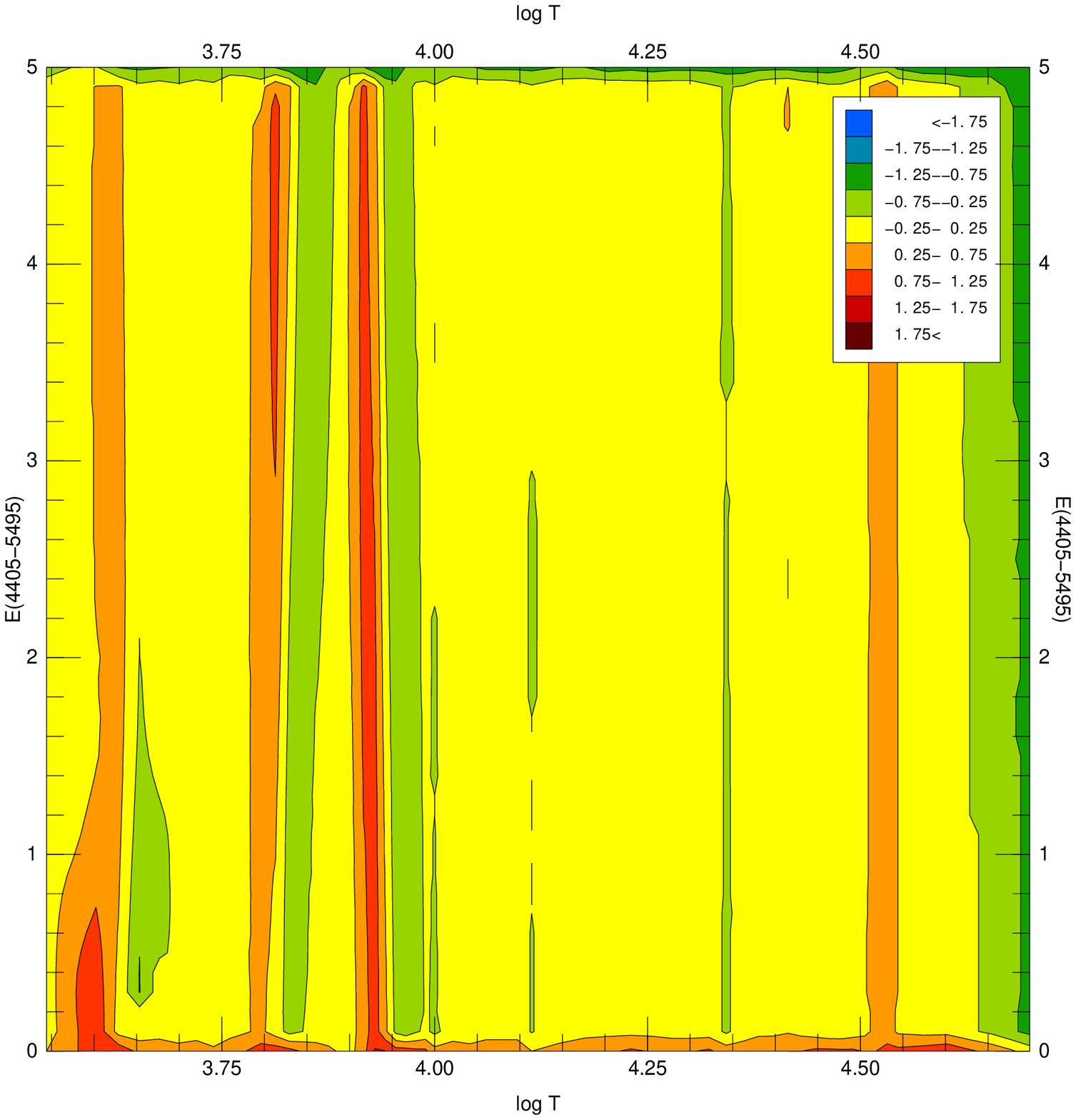}}
\caption{Plots of $d_{E(4405-5495)}$ for sample application 1 without grid extrapolation
in \ebv\ for the $UBV$ (left) and UBVRI (right) cases.}
\label{sample1_noedgext}
\end{figure}

	The importance of extending the grid for some parameters can be seen 
in Fig.~\ref{sample1_noedgext}, where we plot the values of 
$d_{E(4405-5495)}$ for two test runs identical to the ones just described 
with the only exception of not using the grid extension option. We see there
how $|d_{E(4405-5495)}|$ is significantly larger around $\ebv=0.0$ and
$\ebv=5.0$.

\begin{deluxetable}{llrrrcc}
\tablecaption{FITMODEL temperature measurements from Johnson $UBVRI$ photometry.\label{temperature_test}}
\tabletypesize{\scriptsize}
\tablewidth{0pt}
\tablehead{\colhead{Star} & \colhead{Spectral} & \multicolumn{3}{c}{Temperature}                            & \multicolumn{2}{c}{$\chimin/(M-N)$} \\ 
                          & \colhead{type}     & \colhead{Reference} & \colhead{Kurucz} & \colhead{Lejeune} & \colhead{Kurucz} & \colhead{Lejeune}}
\startdata
 HD 36512 &     B0 V & 29800 &        25989$\pm$\phn678 &        25990$\pm$\phn679 &  0.17 &  0.17 \\
 HD 37018 &     B1 V & 26600 &        22704$\pm$\phn721 &        22707$\pm$\phn721 &  2.39 &  2.39 \\
 HD 74280 &     B3 V & 18750 &        16593$\pm$\phn368 &        16595$\pm$\phn369 &  0.12 &  0.12 \\
HD 219688 &     B5 V & 15350 &        13821$\pm$\phn219 &        13830$\pm$\phn215 &  0.15 &  0.16 \\
HD 222661 &   B9.5 V & 10300 &        10109$\pm$\phn206 &        10135$\pm$\phn129 &  0.09 &  0.02 \\
 HD 18331 &     A1 V &  9230 &         9239$\pm$\phn405 &         9097$\pm$\phn528 &  0.78 &  0.50 \\
HD 216956 &     A3 V &  8720 &         8812$\pm$\phn235 &         8595$\pm$\phn402 &  1.45 &  1.41 \\
 HD 26911 &     F3 V &  6700 &      7417$\pm$\phn\phn75 &      6988$\pm$\phn\phn70 &  0.88 &  0.64 \\
 HD 27534 &     F5 V &  6440 &      7226$\pm$\phn\phn95 &      6811$\pm$\phn\phn98 &  1.14 &  0.02 \\
HD 222368 &     F7 V &  6300 &         6707$\pm$\phn243 &         6321$\pm$\phn202 &  0.43 &  0.38 \\
HD 102870 &     F9 V &  6100 &         5906$\pm$\phn227 &         5945$\pm$\phn545 &  2.27 &  0.90 \\
HD 141004 &     G0 V &  6030 &      5895$\pm$\phn\phn76 &         5799$\pm$\phn368 &  0.67 &  0.62 \\
 HD 27836 &     G1 V &  5950 &         6566$\pm$\phn622 &         6524$\pm$\phn418 &  3.02 &  0.81 \\
  HD 1835 &   G2.5 V &  5830 &      5547$\pm$\phn\phn44 &      5399$\pm$\phn\phn37 &  0.55 &  0.32 \\
 HD 20630 &     G5 V &  5770 &      5722$\pm$\phn\phn49 &      5529$\pm$\phn\phn37 &  0.02 &  1.03 \\
 HD 20794 &     G8 V &  5570 &      5776$\pm$\phn\phn52 &      5570$\pm$\phn\phn71 &  0.35 &  0.30 \\
 HD 26965 &   K0.5 V &  5160 &      5363$\pm$\phn\phn36 &      5263$\pm$\phn\phn26 &  0.92 &  0.15 \\
 HD 22049 &     K2 V &  4900 &      5153$\pm$\phn\phn34 &      5122$\pm$\phn\phn21 &  0.22 &  0.44 \\
HD 196795 &     K5 V &  4350 &         4335$\pm$\phn454 &      4021$\pm$\phn\phn42 &  0.53 &  2.15 \\
HD 209290 &   M0.5 V &  3780 &         5303$\pm$\phn143 &         4714$\pm$\phn722 &  3.94 &  3.27 \\
HD 131976 &   M1.5 V &  3650 &        10080$\pm$\phn188 &            5975$\pm$3134 &  3.10 &  0.14 \\
 HD 36395 &   M1.5 V &  3650 &        10083$\pm$\phn168 &            5812$\pm$3070 &  3.83 &  0.19 \\
HD 119850 &     M2 V &  3580 &        10880$\pm$\phn566 &            8924$\pm$3139 &  3.09 &  0.92 \\
\enddata
\end{deluxetable}

	In order to test this application with real data, we selected a sample
of main-sequence stars from the list of MK standard stars by \citet{Garc89} and 
we obtained their $UBVRI$ photometry from the online catalog of 
\citet{Mermetal97}. The stars were selected to have spectral types between B and M
(an example with O stars is analyzed in the next subsection). Results are shown in
Table~\ref{temperature_test}, where the reference temperature-spectral type conversion 
has been obtained from \citet{Bessetal98} for B stars and for the rest of the 
spectral types from \citet{CarrOstl96}. When using real data for this application,
we are not only testing the existence of multiple solutions for a given color 
combination, but also the accuracy of the atmosphere models. For that reason, we
executed CHORIZOS using both Kurucz and Lejeune atmospheres.

\begin{figure}
\centerline{\includegraphics*[width=\linewidth]{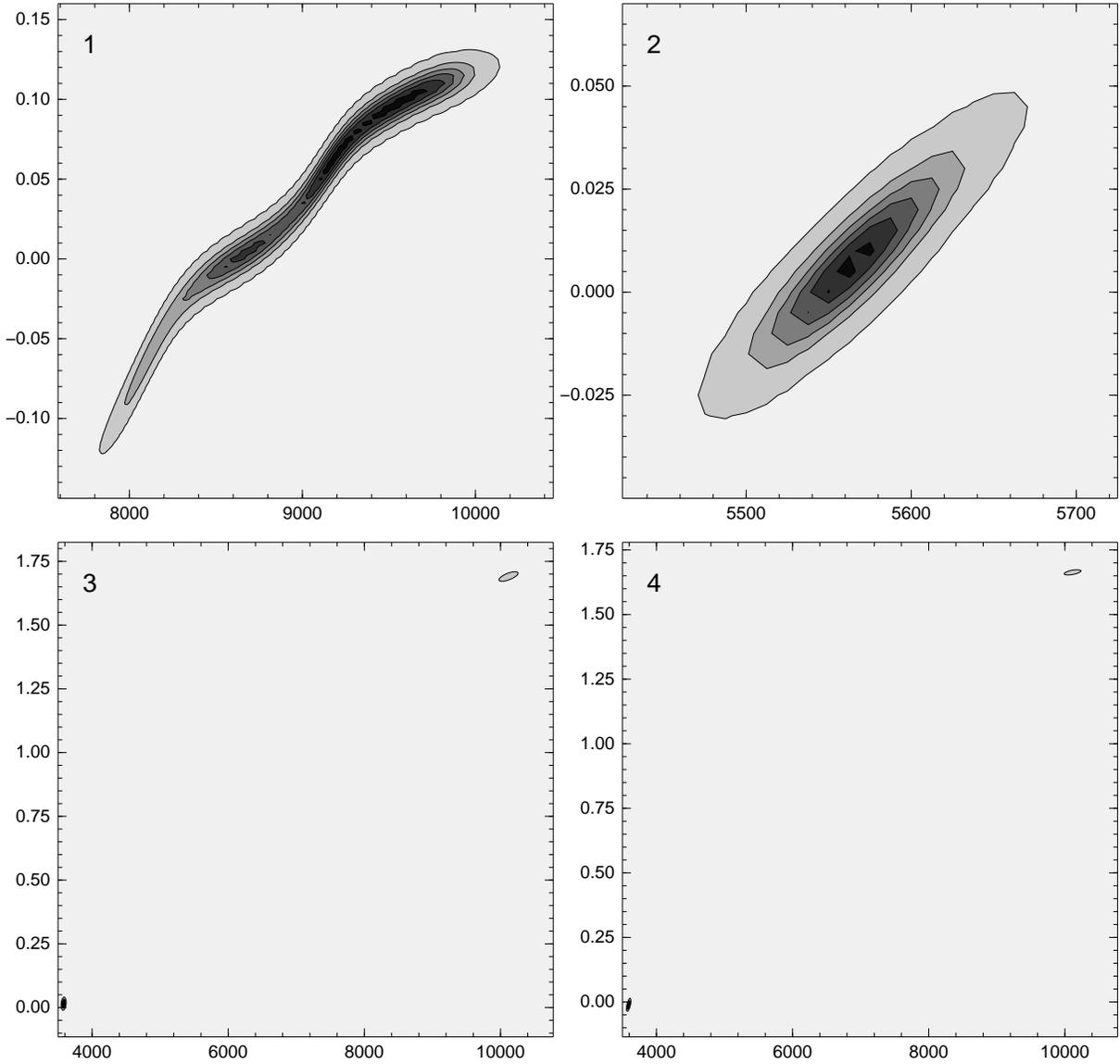}}
\caption{Likelihood contour diagrams for four of the stars in 
Table~\ref{temperature_test} using Lejeune atmospheres. From left to right 
and top to bottom, the cases shown are HD 18331
(A1 V), HD 20794 (G8 V), HD 131976 (M1.5 V), and HD 36395 (M1.5 V). 
Temperature (in K) is plotted in the horizontal axis and \ebv\ in the 
vertical axis.}
\label{sample_stars_lejeune_b}
\end{figure}

\begin{itemize}
  \item For A-K stars, both runs (with Kurucz and Lejeune atmospheres) produce
	good results, once the expected uncertainty in the reference temperature 
	for a given spectral type is included. $\chimin/(M-N)$ values are low
	and a single solution of the correct temperature appears in the likelihood
	plots (see Fig.~\ref{sample_stars_lejeune_b}. The only significant difference 
	between the two atmosphere models for these spectral types takes place for
	F stars, where the Lejeune atmospheres provide a slightly better fit than
	the Kurucz ones.
  \item For B stars, both runs provide essentially identical results but the 
	temperature scale appears to be offset by $\approx 3000$ K for the 
	earliest subtypes. This could be a continuation of the similar effect 
	detected by \citet{GarcBian04} and other authors for O stars using 
	spectroscopic data, as described in the next subsection.
  \item For M stars. the Kurucz run yields bad fits (large values of $\chimin/(M-N)$)
	and clearly incorrect temperatures. This indicates that the Kurucz atmospheres 
	do not provide a good representation for some of the optical colors of M dwarfs, 
	as already pointed out by other authors (see, e.g. \citealt{Lejeetal98}).
        The Lejeune run yields low values of
 	$\chimin/(M-N)$ (except for HD 209290) and acceptable values of $T$ but with 
	large uncertainties. The explanation for the Lejeune results can be seen in the
	two lower plots of Fig.~\ref{sample_stars_lejeune_b}: two solutions are 
	present, the correct one around 3\,650 K and another one around 10\, 000 K. 
	This implies that the Lejeune atmospheres provide a better representation of 
	the colors of M dwarfs but that $UBVRI$ photometry alone is not enough to 
	distinguish between M stars and reddened late-B stars.
\end{itemize}

	The conclusion is that the addition of $RI$ data to $UBV$ photometry is
very useful in constraining the temperature and extinction of stars (under
certain assumptions such as knowledge of the extinction law and the luminosity
type) and that CHORIZOS can be successfully used to derive those parameters. 
A possible application of the code would be to automatically generate temperatures
and extinctions from photometric surveys such as the one planned with the GAIA 
mission.

\subsection{Measuring optical-IR extinction laws}

	As a second example, we use Johnson $UBV$ and 2MASS $JHK_s$ to determine 
the extinction and extinction law experienced by an early-type star. We select as 
input a 35\,000 K, $\log g = 5.0$, solar metallicity Kurucz atmosphere model with
uncertainties of 0.01 magnitudes in each of the six filters. The Kurucz atmosphere 
is extinguished from \ebv = 0.0 to \ebv = 5.0 using \citet{Cardetal89} laws from \rv = 2.0
to \rv = 6.0.

\begin{figure}
\centerline{\includegraphics*[width=0.48\linewidth]{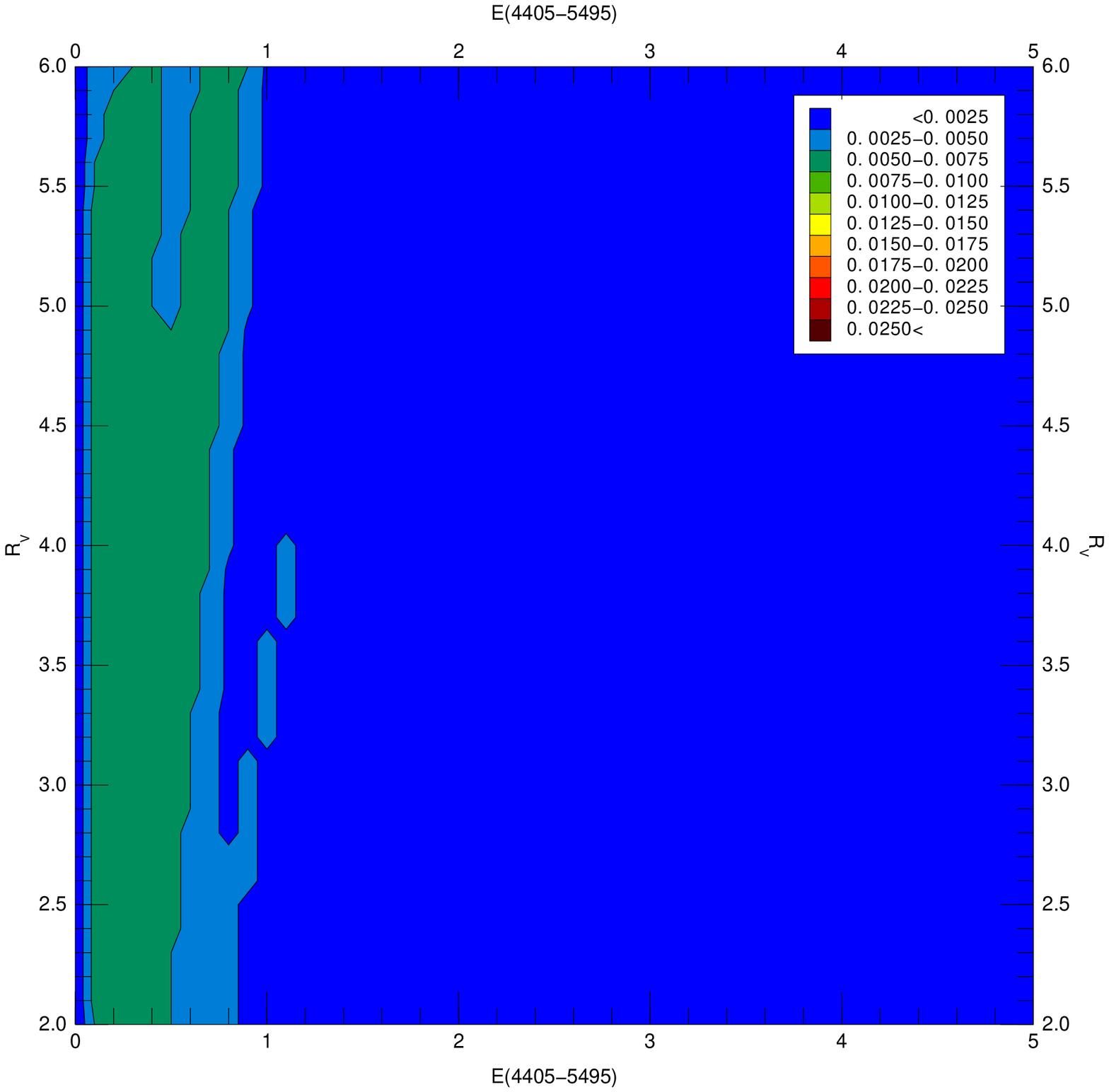}
          \ \includegraphics*[width=0.48\linewidth]{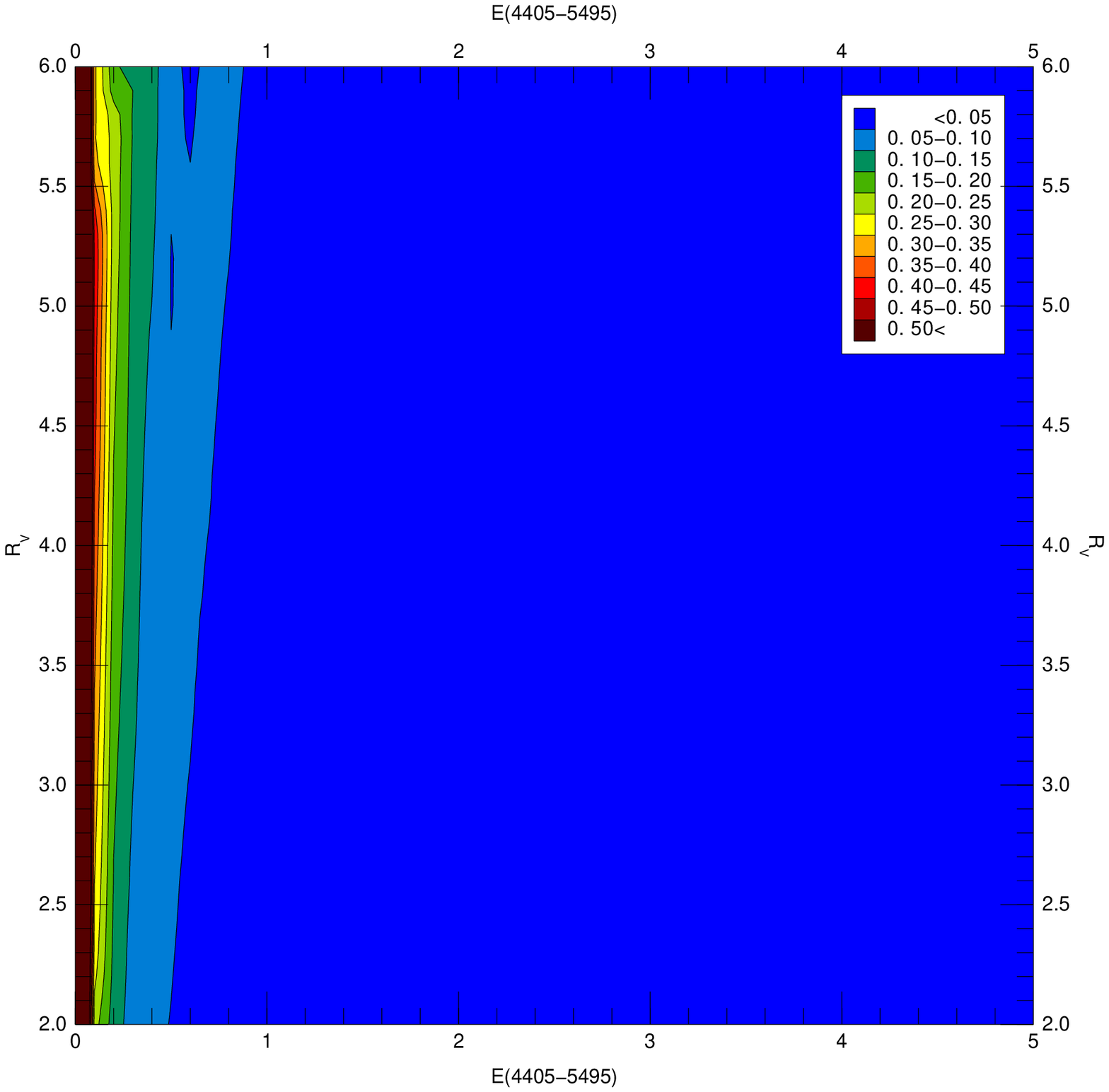}}
\centerline{\includegraphics*[width=0.48\linewidth]{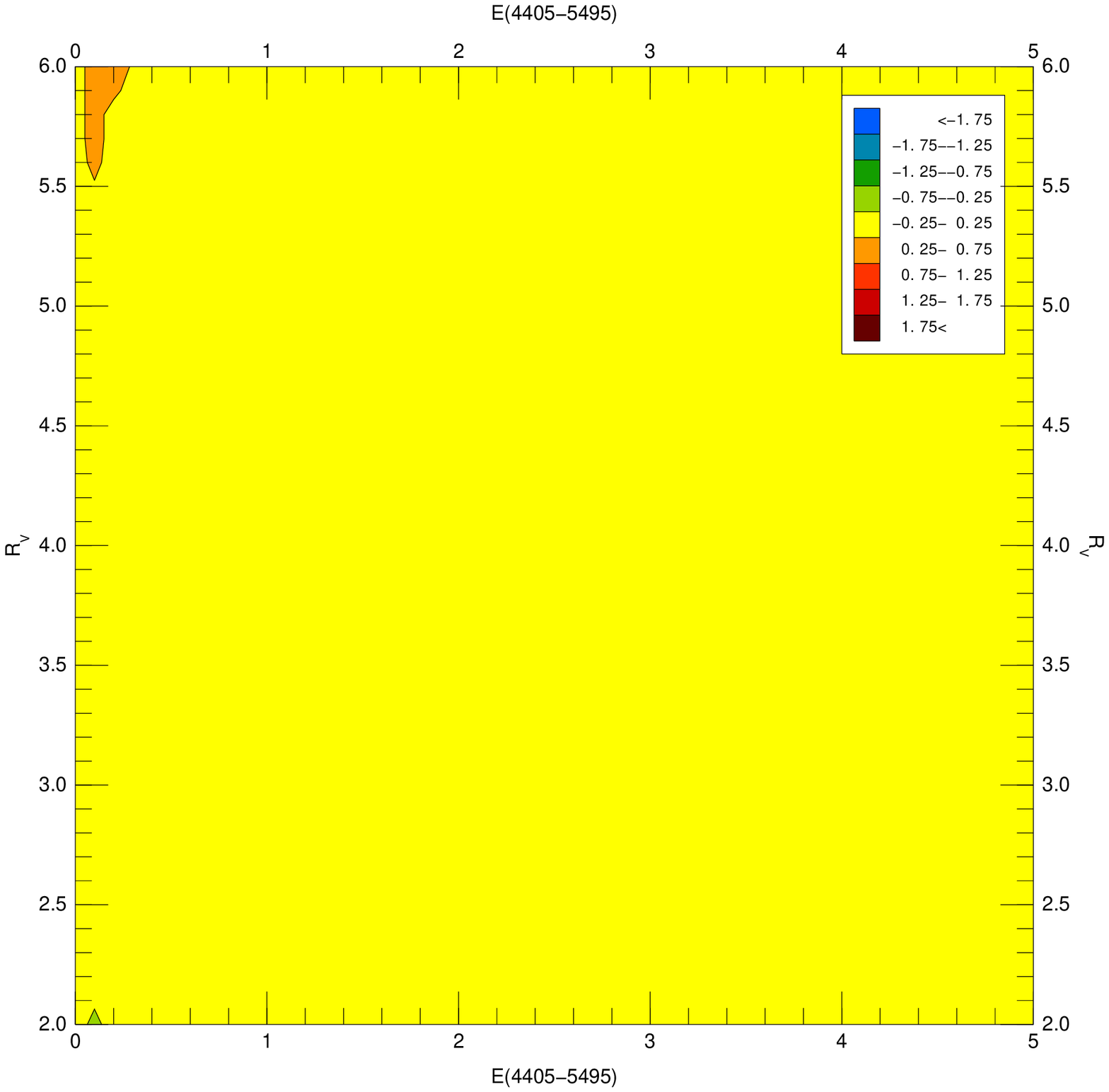}
          \ \includegraphics*[width=0.48\linewidth]{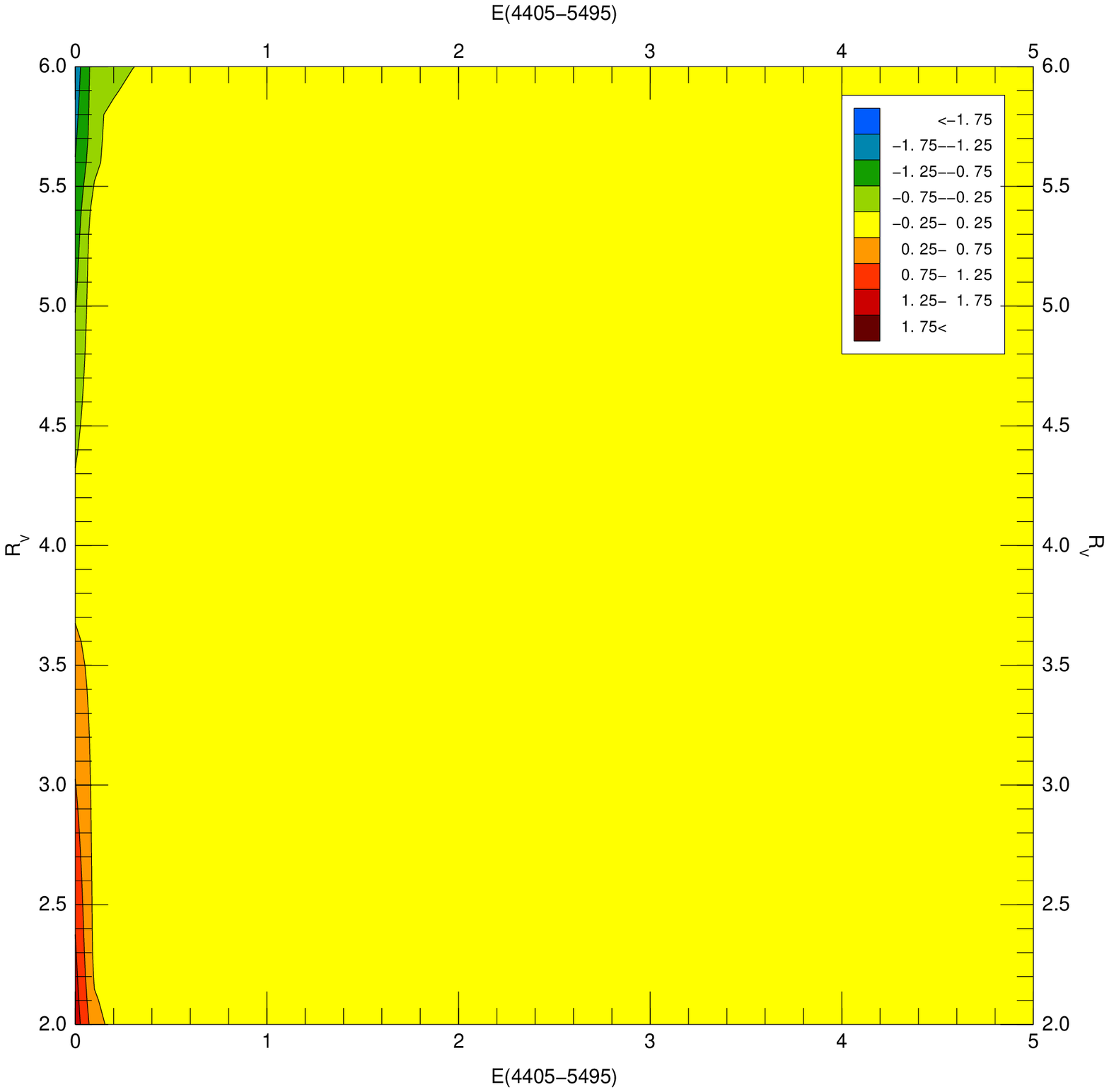}}
\caption{(top) Plots of $\sigma_{E(4405-5495)}$ (left) and $\sigma_{R_{5495}}$ (right) 
for sample application 2 (main-sequence 35\,000 K Kurucz model) as a function of \ebv\ 
and \rv\ calculated for $UBVJHK$ photometry with uncertainties of 0.01 magnitudes for 
each filter and restricting the model temperature to its true value. (bottom) Plots of
$d_{E(4405-5495)}$ (left) and $d_{R_{5495}}$ (right) corresponding to the same case.}
\label{sample2}
\end{figure}

	We first assume that we have an accurate spectral type for the star and, 
therefore, that we know a priori the temperature and gravity of the star. We do this 
by constraining the temperature and gravity in CHORIZOS to a single value (the true one).
Grid extension is used for both \ebv\ and \rv.
Results for this $M=5$, $N=2$ case are shown in Fig.~\ref{sample2}. 
The two lower plots demonstrate that $UBVJHK_s$
is an adequate choice of filters to measure the extinction and extinction law experienced
by hot stars. The values of both normalized distances are very close to 0.0 everywhere 
with the only exception of the region around \ebv  = 0 for $d_{R_{5495}}$. The latter is 
an expected behavior, since for low values of the reddening all extinction laws produce 
similar results and for \ebv = 0.0 they are strictly degenerate. This is evidenced in the upper 
right plot of Fig.~\ref{sample2}: the value of $\sigma_{R_{5495}}$ is kept lower than 0.10
for $\ebv > 0.4$ but increases rapidly as we approach $\ebv = 0.0$, where the lack of 
information on the extinction law provided by the data manifests itself in large values of
$\sigma_{R_{5495}}$. Note, however, that $\sigma_{E(4405-5495)}$ is not strongly affected 
by this, since its value is kept below 0.0075 everywhere.

\begin{figure}
\centerline{\includegraphics*[width=0.48\linewidth]{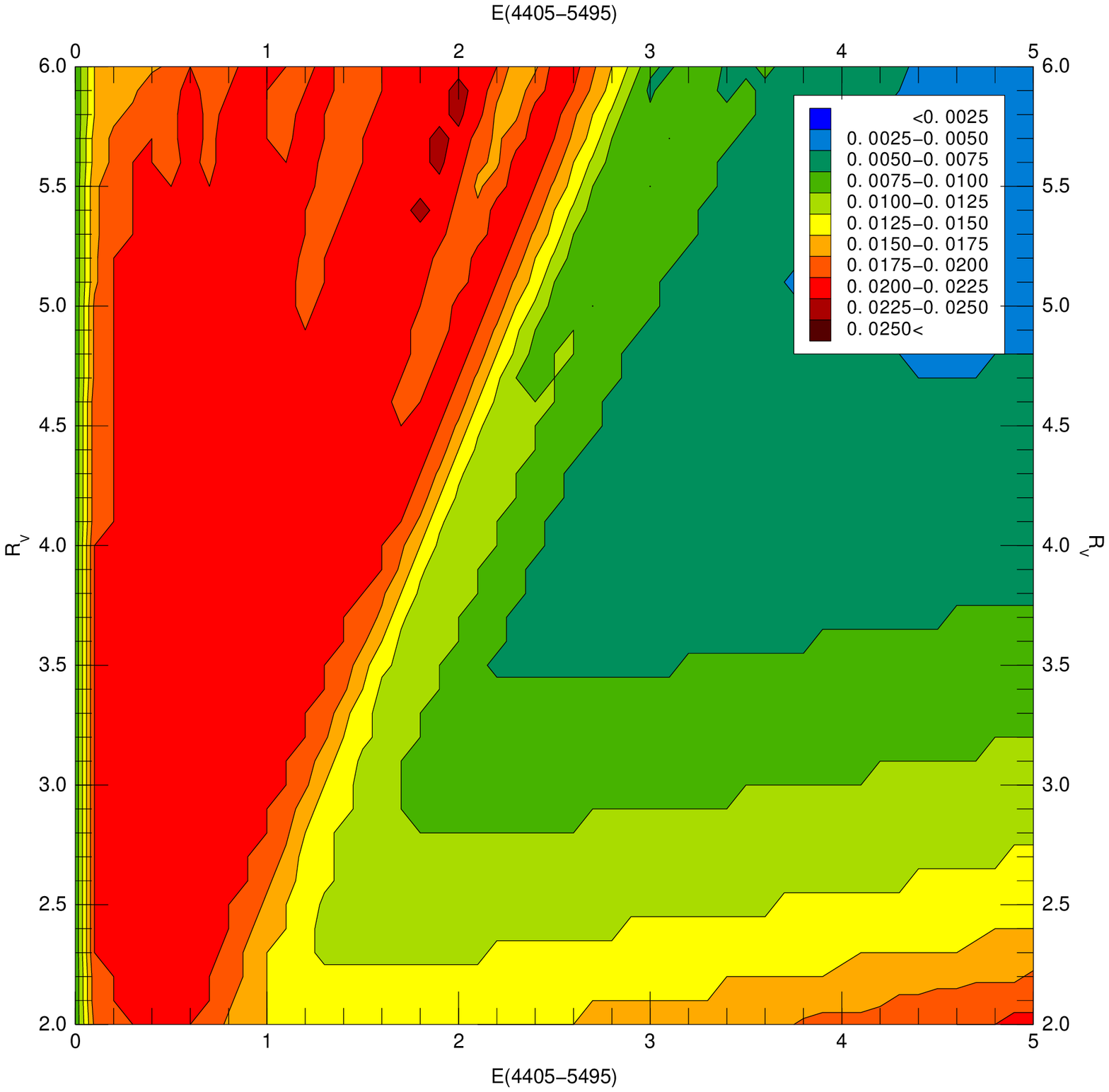}
          \ \includegraphics*[width=0.48\linewidth]{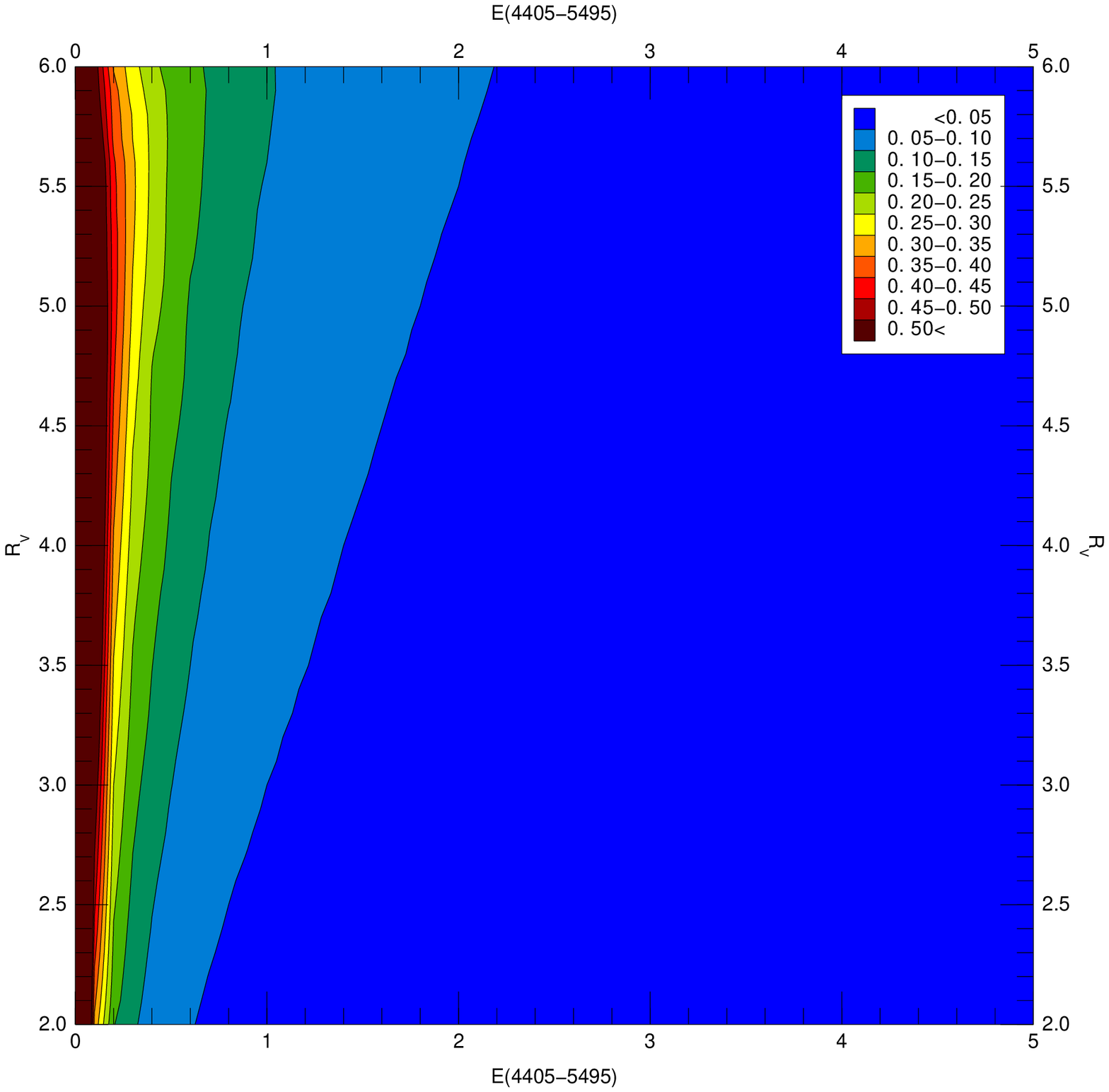}}
\centerline{\includegraphics*[width=0.48\linewidth]{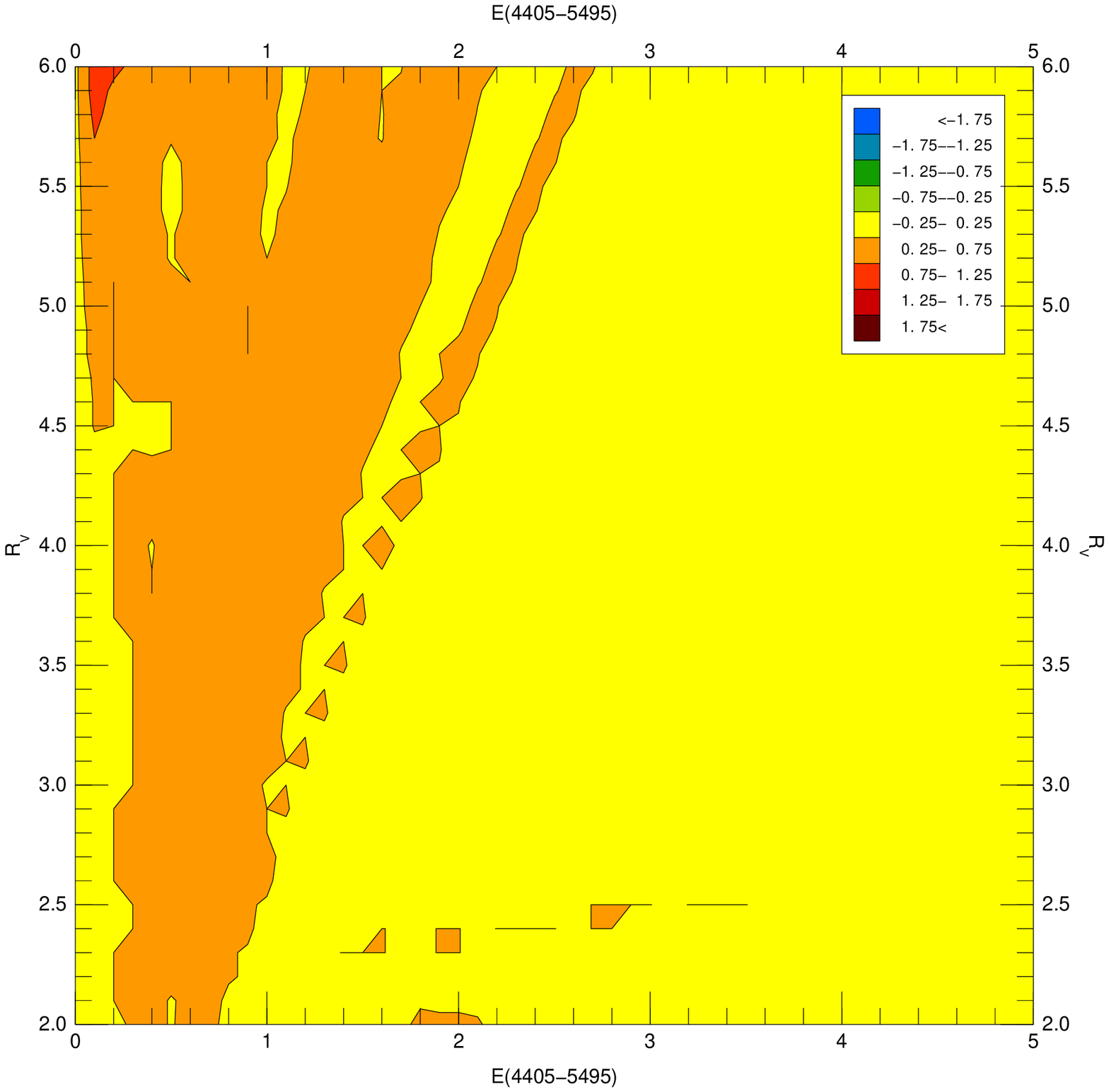}
          \ \includegraphics*[width=0.48\linewidth]{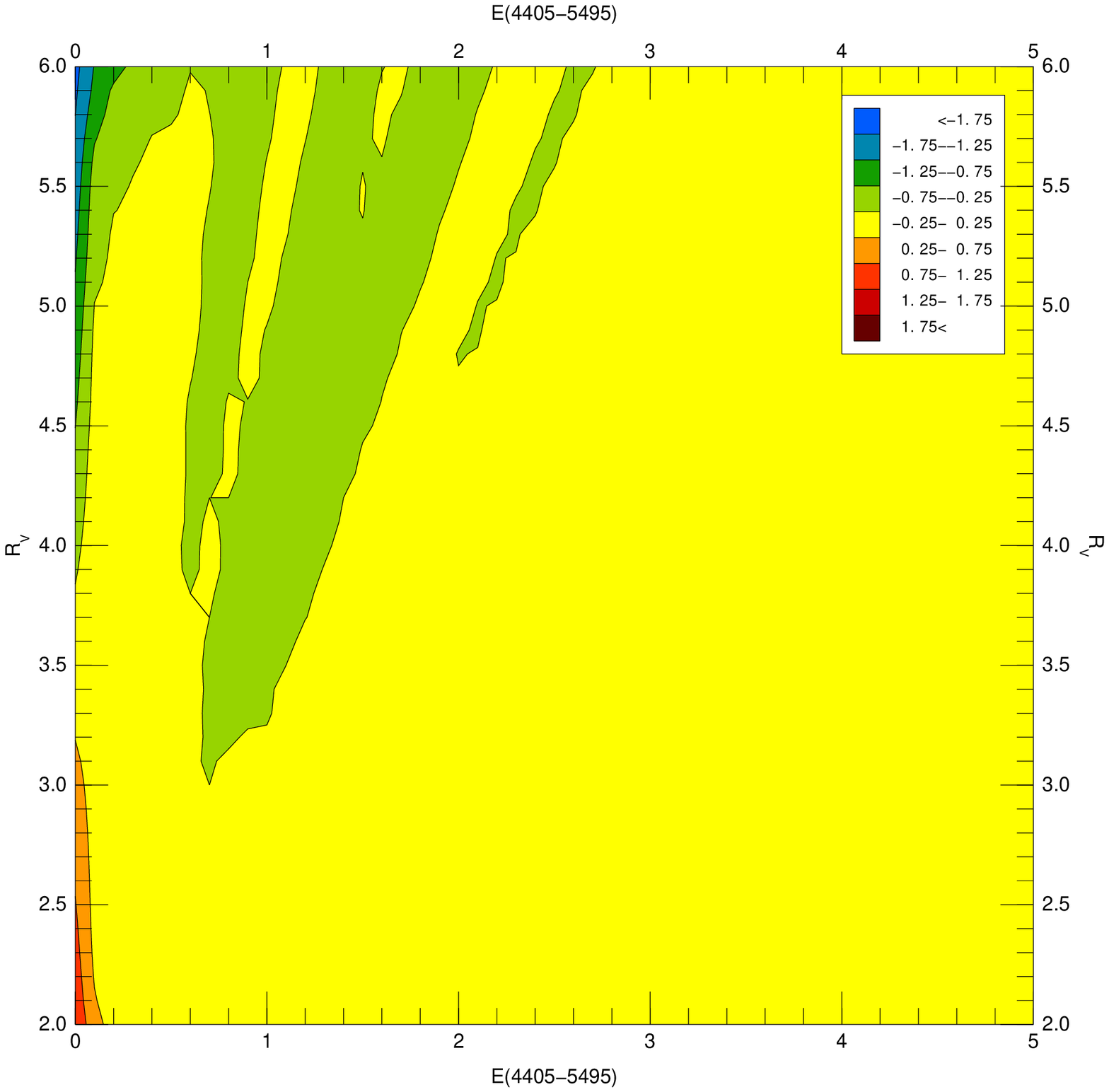}}
\caption{Same as Fig.~\ref{sample2} but without placing constraints on the model 
temperature.}
\label{sample2_freet}
\end{figure}

	Suppose now that we know that the star is of early type but we cannot constrain its 
temperature farther than that due to the lack of an accurate spectral type. We can
simulate such a case in CHORIZOS by leaving the temperature unconstrained and selecting
main-sequence Kurucz models. Results for this $M=5$, $N=3$ case are shown in 
Fig.~\ref{sample2_freet}. We see in the two lower plots there that the normalized distances
are slightly worse than in the previous case, but still within acceptable ranges. The 
degeneracy in \rv\ is still present for \ebv = 0.0, as expected, but the information in the 
photometry is accurate enough to yield good estimates of both \ebv\ and \rv. The loss of 
information caused by the unconstrained temperature only translates into larger uncertainties
in the measured quantities, but even those are smaller than 0.03 anywhere for 
$\sigma_{E(4405-5495)}$ and lower than 0.25 for $\sigma_{R_{5495}}$ for $\ebv > 0.4$.

\begin{deluxetable}{llcccccccccccc}
\tablecaption{Comparison between FITMODEL and \citet{Cardetal89} results for the reddening and extinction law
of 4 O stars.\label{extinction_test}}
\rotate
\tabletypesize{\scriptsize}
\tablewidth{0pt}
\tablehead{\colhead{Star} & \colhead{Spectral} & \multicolumn{2}{c}{\citet{Cardetal89}} & \multicolumn{5}{c}{Constrained $T$ and $\log g$ ($M=5$, $N=2$)} & \multicolumn{5}{c}{Unconstrained $T$ and $\log g$ ($M=5$, $N=4$)} \\
                          & \colhead{type}     & $E(B-V)$ & $R_V$                       & $T$ (K) & $\log g$ & \ebv & \rv & $\frac{\chimin}{M-N}$                         & $T$ (K) & $\log g$ & \ebv & \rv & $\frac{\chimin}{M-N}$  }
\startdata
HD 46202        & O9 V           & 0.47 & 3.12 & 32\,500 & 4.00 & 0.478$\pm$0.013 & 3.05$\pm$0.12 & 1.70 & 28\,900$\pm$1\,800 & 4.13$\pm$0.46 & 0.448$\pm$0.020 & 3.14$\pm$0.14 & 0.57 \\
HD 73882        & O8.5 V ((n))   & 0.72 & 3.39 & 32\,500 & 4.00 & 0.704$\pm$0.017 & 3.42$\pm$0.10 & 3.78 & 28\,600$\pm$1\,900 & 4.21$\pm$0.43 & 0.679$\pm$0.021 & 3.44$\pm$0.10 & 5.64 \\
HD 229196       & O6 III (n)(f)  & 1.22 & 3.12 & 37\,500 & 3.75 & 1.221$\pm$0.014 & 3.16$\pm$0.05 & 0.47 & 37\,400$\pm$3\,400 & 3.88$\pm$0.56 & 1.214$\pm$0.028 & 3.16$\pm$0.06 & 1.32 \\
CPD $-$59 2600  & O6 V ((f))     & 0.53 & 4.17 & 37\,500 & 4.00 & 0.507$\pm$0.023 & 4.13$\pm$0.21 & 1.14 & 36\,500$\pm$3\,500 & 3.90$\pm$0.56 & 0.492$\pm$0.033 & 4.14$\pm$0.23 & 2.17 \\
\enddata
\end{deluxetable}

	We also tested the measurement of optical-IR extinction laws with CHORIZOS using
real data from the literature. We selected from
the sample used by \citet{Cardetal89} to derive their extinction law the four stars present in
the Galactic O star catalog of \citet{Maizetal04b} which have (a) $UBVJHK_s$ data in the 
catalog and (b) values of $E(B-V)$ measured by \citet{Cardetal89} greater than 0.45. CHORIZOS 
was run using the $UBVJHK_s$ photometry ($M=5$) and TLUSTY atmospheres twice, first 
constraining the temperatures and gravities to fixed values ($N=2$) and then leaving them 
unconstrained ($N=4$). The values for the temperatures and gravities were derived from the
spectral types by using a scale intermediate between the ones proposed by \citet{Vaccetal96} 
and \citet{GarcBian04} and then selecting the closest TLUSTY model. Results are
shown in Table~\ref{extinction_test}.

\begin{itemize}
  \item CHORIZOS results for \rv\ are in excellent agreement with those of \citet{Cardetal89}.
	That article does not provide error estimates but our results are always within one
	sigma and are also small enough for the output to be meaningful.
  \item CHORIZOS results for \ebv\ are also in very good agreement with the reference values, 
	though the unconstrained results are in all cases lower than the ones provided by 
	\citet{Cardetal89} (but always within two sigmas). The likely origin of this minor 
	difference is the use of values between $-0.30$ and $-0.32$ for the $(B-V)_0$ colors
	of O stars by those authors. TLUSTY atmospheres predict redder colors by $\approx 0.02$
	magnitudes. Once that difference is included, the agreement is excellent. 
  \item The \chimin\ values indicate very good fits in all cases except for HD 73882, which is
	still acceptable.
  \item Assuming that no biases are present (e.g. systematic errors in the TLUSTY atmospheric models), 
	the unconstrained results favor the lower temperature scale of \citealt{GarcBian04} 
	(O4 dwarfs around 40\, 000 K and O6 dwarfs around 32\,000-35\,000 K, implying a 
	boundary between O and B dwarfs below 30\,000 K) over that of \citealt{Vaccetal96}
 	(with O4 dwarfs close to 50\,000 K and a boundary between O and B dwarfs at 34\,000 K).
\end{itemize}

	In summary, CHORIZOS can be used to measure reddenings and extinction laws with good
precision, even when accurate spectral types are not available.

\subsection{What precision is required to measure gravity for O stars with optical photometry 
alone?}

	For our third example we want to investigate the possibility of using Str\"omgren
photometry to measure the surface gravity of O stars of unknown temperature and extinction
(but with a known extinction law). This is obviously a difficult task, since O-star 
optical colors are quasi-degenerate in temperature and even more so in gravity. Our goal will
be to determine what kind of photometric precision would be required and to assess whether
such an accuracy is attainable.

	We use TLUSTY atmospheres with solar metallicity, $T = 35\,000$ K, and 
$\log g$ between 3.25 and 4.75 observed with Str\"omgren $ubvy$ photometry. The extinction law 
is restricted to be of \citet{Cardetal89} type with $\rv = 3.1$ but no constraints are placed 
on the possible values of $T$, $\log g$, or \ebv, yielding an $M=3$, $N=3$ case. Grid extension is
used for \ebv\ but not for $T$ or $\log g$ due to the incompleteness of the model grid in 
these two last parameters \citep{LanzHube03}.

\begin{figure}
\centerline{\includegraphics*[width=0.48\linewidth]{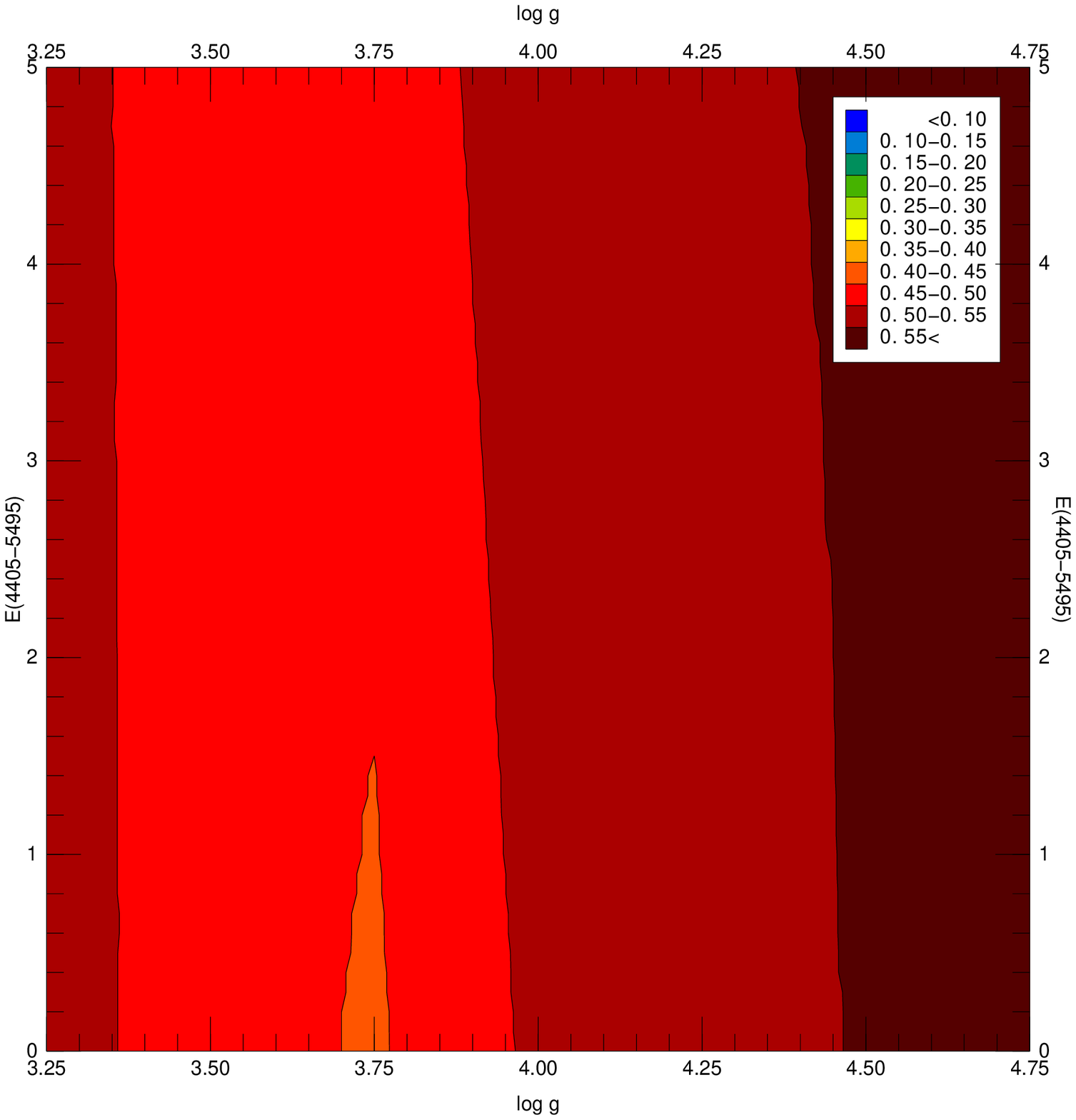}
          \ \includegraphics*[width=0.48\linewidth]{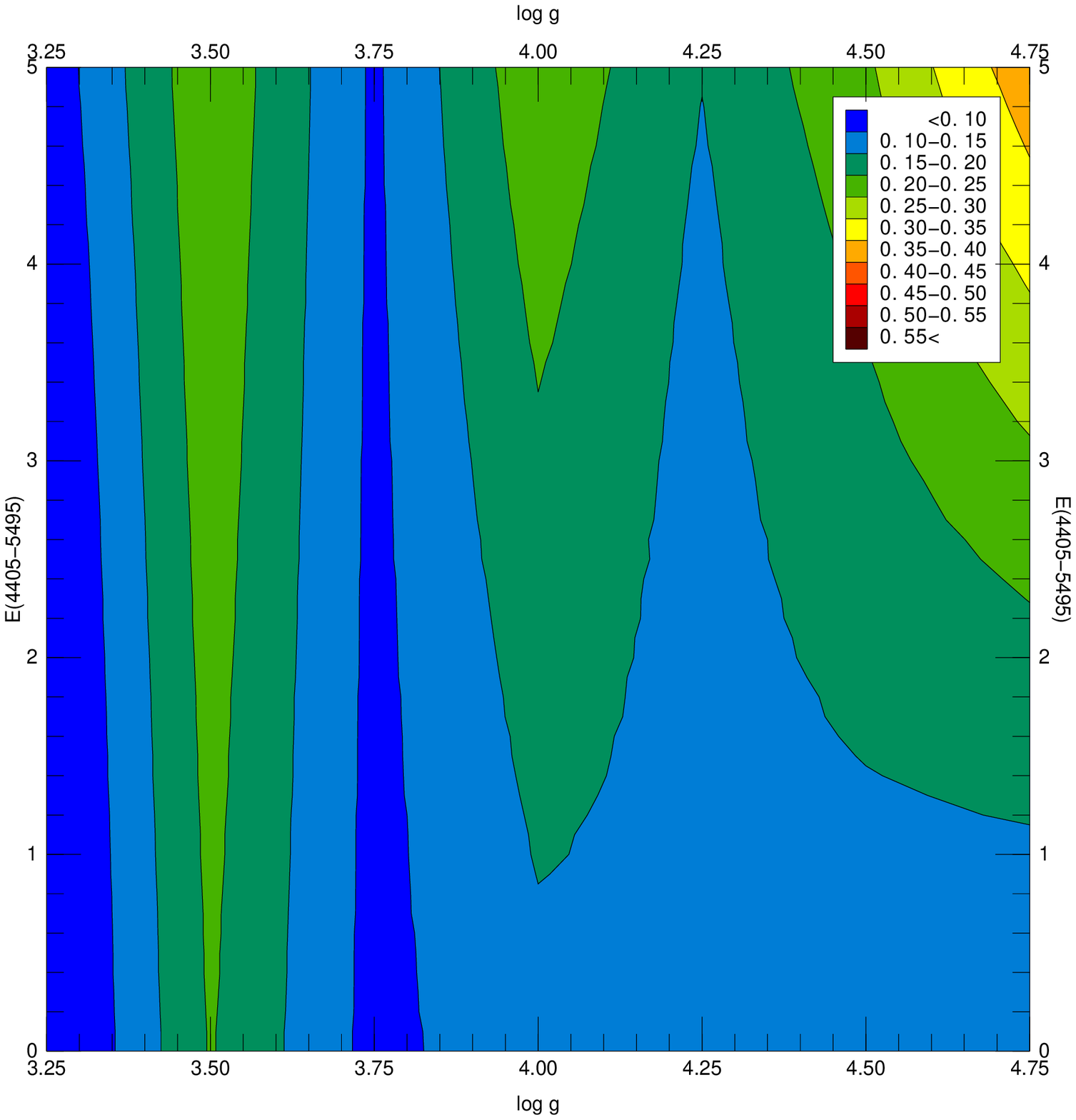}}
\centerline{\includegraphics*[width=0.48\linewidth]{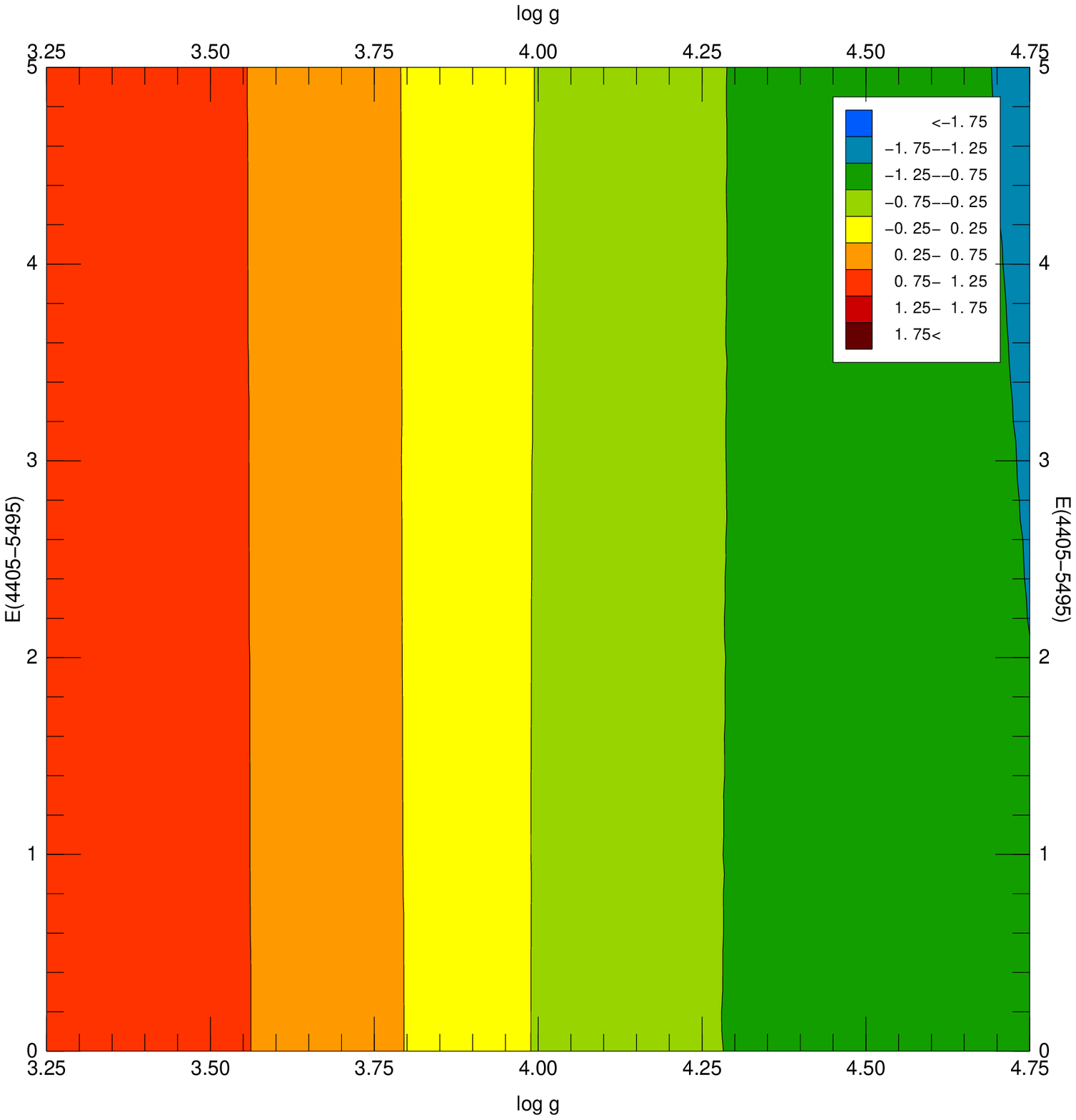}
          \ \includegraphics*[width=0.48\linewidth]{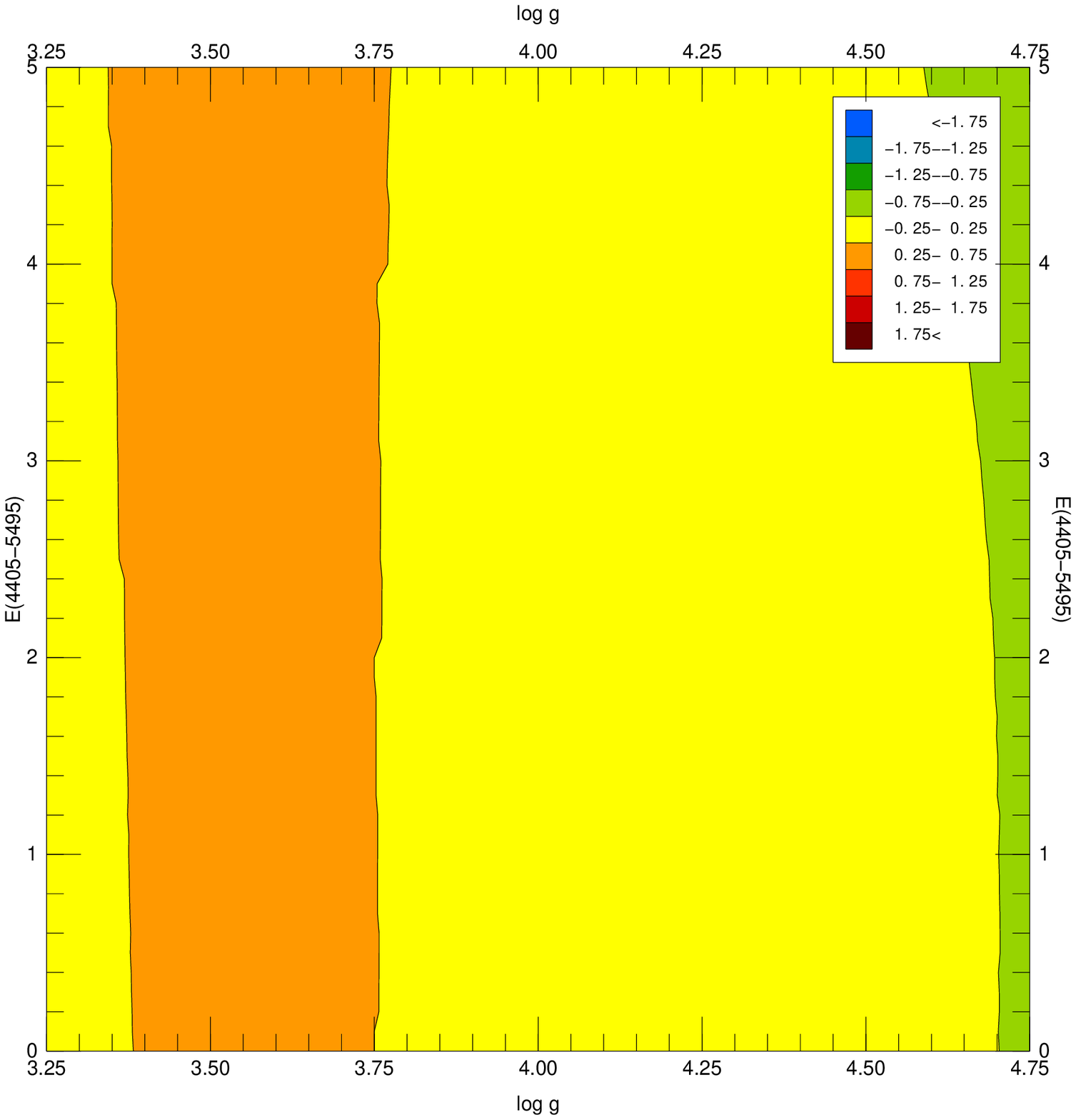}}
\caption{(top) Plots of $\sigma_{\log g}$ for sample application 3 (TLUSTY solar metallicity
35\,000 K models) as a function of $\log g$ and \ebv\ calculated for Str\"omgren photometry
with uncertainties of 0.003 (left) and 0.001 (right) magnitudes for each filter.
(bottom) Plots of $d_{\log g}$ corresponding to the same cases.}
\label{sample3}
\end{figure}

	In a first run, values of 0.003 were used for the uncertainties in the measured magnitudes.
Results for $\sigma_{\log g}$ and $d_{\log g}$ are shown in the left panels of Fig.~\ref{sample3}.
We see there that the supplied photometry does not yield enough information to measure gravities.
The high values of both $d_{\log g}$ and $\sigma_{\log g}$ are characteristic of an almost constant 
output result of $\approx 4.0\pm 0.5$ for $\log g$ independent of the input values for the
gravity and reddening.

	In a second run we use values of 0.001 for the uncertainties. Results are shown in the
right panels of Fig.\ref{sample3}. Now the values for $d_{\log g}$ and $\sigma_{\log g}$ show that
some discrimination is possible among gravities, with typical uncertainties in $\log g$ around 
$0.10-0.20$, which is enough to differentiate between main sequence stars and supergiants.

	A hypothetical observer should now ask him/herself: Am I convinced that the atmosphere models
are correct to within 0.001 magnitudes? Can I calibrate the photometry to levels good enough to
accurately measure differences of 1 milimagnitude? At the current level of knowledge and technology 
the answers to both questions are likely to be no, hence it can be deduced that at the present time
optical photometry alone cannot be used to accurately measure O-star gravities. However, future 
improvements in atmosphere modeling and photometric accuracies may change the situation.

\section{Summary}

	CHORIZOS is a multi-purpose $\chi^2$-minimization SED-fitting code that 
can be applied to either photometric or spectrophotometric data. In this
article we have described the techniques it employs and applied it to several
astronomical examples. At the time of this writing, a
beta version of the program with a limited number of SED models and 78 filter
passbands is available from {\tt http://www.stsci.edu/\~{}jmaiz}. In the
future, a full version will be made available which will include user-defined
filter sets or wavelength ranges as well as the possibility of adding other
SED models.

	Besides the obvious application of selecting the model(s) which are
compatible with a given set of observed data, CHORIZOS can be used for a 
number of other astronomical applications. It can be utilized to select the 
optimum choice of filters and minimum S/N requires when planning an 
observation, to test atmospheric models and extinction laws, or to calibrate
the zero points of a filter system, to name a few.

\acknowledgments

I would like to thank Leonardo \'Ubeda for his help with the testing of the
code, an anonymous referee for his/her helpful comments, and Rodolfo Xeneize
Barb\'a for his christening suggestions.

\bibliographystyle{apj}
\bibliography{general}

\end{document}